# A Mathematical Model of Tripartite Synapse: Astrocyte Induced Synaptic Plasticity

Shivendra Tewari[†]
*shivendra.tewari@gmail.com*
*Systems Science and Informatics Unit*
*Indian Statistical Institute*
*8th Mile, Mysore Road*
*Bangalore 560059, India*

Kaushik Majumdar
*kmajumdar@isibang.ac.in*
*Systems Science and Informatics Unit*
*Indian Statistical Institute*
*8th Mile, Mysore Road*
*Bangalore 560059, India*

In this paper we present a biologically detailed mathematical model of tripartite synapses, where astrocytes modulate short-term synaptic plasticity. The model consists of a pre-synaptic bouton, a post-synaptic dendritic spine-head, a synaptic cleft and a peri-synaptic astrocyte controlling $Ca^{2+}$ dynamics inside the synaptic bouton. This in turn controls glutamate release dynamics in the cleft. As a consequence of this, glutamate concentration in the cleft has been modeled, in which glutamate reuptake by astrocytes has also been incorporated. Finally, dendritic spine-head dynamics has been modeled. As an application, this model clearly shows synaptic potentiation in the hippocampal region, i.e., astrocyte $Ca^{2+}$ mediates synaptic plasticity, which is in conformity with the majority of the recent findings (Perea & Araque, 2007; Henneberger et al., 2010; Navarrete et al., 2012).

## 1 Introduction

One of the most significant challenges in neuroscience is to identify the cellular and molecular processes that underlie learning and memory formation (Lynch, 2004). Cajal originally hypothesized that information storage relies on changes in strength of synaptic connections between neurons that are active (Cajal, 1913). Hebb supported this hypothesis and proposed that if two neurons are active at the same time, the synaptic efficiency of the appropriate synapse will be strengthened (Hebb, 1949). Synaptic transmission is a dynamic process. Post-synaptic responses wax and wane as pre-synaptic activity evolves. Forms of synaptic enhancement, such as facilitation, augmentation, and post-tetanic potentiation, are usually attributed to effects of a residual elevation in pre-synaptic $Ca^{2+}$ concentration ($[Ca^{2+}]$), acting on one or more molecular targets that appear to be distinct from the secretory trigger responsible for fast exocytosis and phasic release of transmitter to single action potential (Zucker & Regehr, 2002). It is now well established that the astrocytic mGluR detects synaptic activity and responds via activation of the calcium-induced calcium release pathway, leading to elevated $Ca^{2+}$ levels. The spread of these levels within micro-domain of one cell can coordinate the activity of disparate synapses that are associated with the same micro-domain (Perea & Araque, 2002). The notion of tripartite synapse consisting of pre-synaptic neuron, post-synaptic neuron and astrocyte has taken a firm root in experimental (Araque, et al., 1999; Newman, 2003; Perea & Araque, 2007) as well as theoretical neuroscience (Nadkarni & Jung, 2003; Volman et al., 2007; Nadkarni, et al., 2008). Astrocytes play crucial roles in the control of Hebbian plasticity (Fellin, 2009).

---

[†] Shivendra Tewari's present address is Biotechnology & Bioengineering Center and Department of Physiology, Medical College of Wisconsin, 8701 Watertown Plank Road, Milwaukee, WI 53226, USA.

There is a recent report, that at least in the hippocampus, astrocyte $Ca^{2+}$ signaling does not modulate short-term or long-term synaptic plasticity (Agulhon, et al., 2010). However evidences of astrocytic modulation of synaptic plasticity are more abundant including in hippocampus (Vernadakis, 1996; Haydon, 2001; Yang et al., 2003; Andersson, 2010; Henneberger, et al., 2010). Neuronal activities can trigger $Ca^{2+}$ elevations in astrocytes (Porter & McCarthy, 1996; Fellin, 2009) leading to concentration increase in adjacent glial cells including astrocytes, which expresses a variety of receptors (Newman, 2003). These activated receptors increase astrocyte $[Ca^{2+}]$, and release transmitters, including glutamate, D-serine, ATP (Parpura et al., 1994; Henneberger et al., 2010) etc. The released gliotransmitters feed-back onto the pre-synaptic terminal either to enhance or to depress further release of neurotransmitter (Newman, 2003; Navarrete & Araque, 2010) including glutamate, which is mediated by $Ca^{2+}$ concentration in the pre-synaptic terminal. It is worthy to note that $Ca^{2+}$ elevations are both necessary and sufficient to evoke glutamate release from astrocytes (Haydon, 2001). On the other hand short-term synaptic depression is caused by depletion of the releasable vesicle pool due to recent release in response to pre-synaptic action potential (Wu & Borst, 1999). This entire chain of $Ca^{2+}$ mediated pre-synaptic activity consisting of both short-term enhancement (STE) and short-term depression (STD) can be called short-term synaptic plasticity or simply short-term plasticity (STP).

Synaptic plasticity occurs at many time scales. Usually long-term plasticity (LTP) happens at a time scale of 30 minutes or more and STP takes less than that (p – 311, Koch, 1999). Within the ambit of STP, STE has been more widely studied than the STD. A quantitative definition of STE has been proposed in (Fisher et al., 1997). STE has been divided into four different temporal regimes, namely fast-decaying facilitation (tens of milliseconds), slow-decaying facilitation (hundreds of milliseconds), augmentation (seconds) and post-tetanic potentiation (minutes) (Fisher et al., 1997).

STP is thought to provide a biological mechanism for on-line information processing in the central nervous system (Fisher et al., 1997) and therefore could be the key to the formation of working memory and intelligent behavior. A computational model of how cellular and molecular dynamics give rise to the STP in the synapses (particularly in the synapses of the hippocampus and the prefrontal cortex) can be quite useful in understanding intelligent behavior.

In this paper, we present a computational model of astrocyte mediated synaptic potentiation in a tripartite synapse. The present model is based on experimental work of Perea & Araque (2007) where they used immature wistar rats for hippocampal slice preparations. Primarily there are just two models (Nadkarni et al., 2008; Volman et al., 2007) shedding light over the molecular aspects of astrocyte mediated synaptic potentiation, where a lot of important details were omitted or were modeled hypothetically (see Table 1).

Table 1: A Comparison among Nadkarni et al (2008) model, Volman et al (2007) model, and the proposed model

| Signaling Processes Modeled | Volman et al., 2007 | Nadkarni et al., 2008 | This Paper |
|---|---|---|---|
| Bouton $Ca^{2+}$ | No | Yes | Yes |
| Bouton $IP_3$ | No | No | Yes |
| Synaptic Vesicle / Glutamate | Yes / No | Yes / No | Yes / Yes |
| Astrocytic $Ca^{2+}$ | Yes | Yes | Yes |
| Astrocytic $IP_3$ | Yes | Yes | Yes |
| Extra-synaptic Vesicle / Glutamate | No | No | Yes / Yes |
| Post-Synaptic Current / Potential | Yes / No | Yes / No | Yes / Yes |

The computational model proposed here makes use of different detailed biophysical models highlighting specific aspects of astrocyte-neuron signaling. The following steps have been followed in simulation of our model. (1) Pre-synaptic action potential train has been generated using the HH model (Hodgkin & Huxley, 1952). (2) $Ca^{2+}$ concentration elevation in the pre-synaptic bouton incorporating fast (using single protein properties (Erler et al., 2004)) and slow (using modified Li-Rinzel model (Li & Rinzel, 1994)) $Ca^{2+}$ influx. (3) Glutamate release in the synaptic cleft as a two step process (using Bollman et al., (2000) for $Ca^{2+}$ binding to synaptic vesicle sensor and, Tsodyks & Markram (1999) for synaptic vesicle fusion and recycling). (4a) Glutamate modulated enhancement of astrocytic $Ca^{2+}$ (using astrocyte specific G-Chi model (De Pitta et al., 2009)). (4b) Glutamate mediated excitatory post-synaptic current (using Destexhe et al (1999)) and potential (using Tsodyks & Markram (1997)). (5) Extra-synaptic glutamate elevation is also modeled as a two-step process (using modified Bertram model (Bertram et al., 1996) to fit Synaptic-Like Micro-vesicle (SLMV) release probability determined recently (Malarkey & Parpura, 2011) and, Tsodyks & Markram (1997) for SLMV fusion and recycling). The motivations and consequences of the specific models chosen have been explained in appropriate places.

We observed an increase in average neurotransmitter release probability, Pr, after astrocyte became active (before: 0.25; after: 0.35) which is in close conformity with the experimental observation (before: 0.24; after: 0.33) of Perea & Araque (2007). On measuring the windowed average amplitude of the excitatory post-synaptic current (EPSC) we could observe up to 250% increase from pre-astrocytic activities to the post-astrocytic activities, which decayed with a time constant of 10 to 12 seconds. This signifies augmentation (Fisher et al., 1997; Koch, 1999).

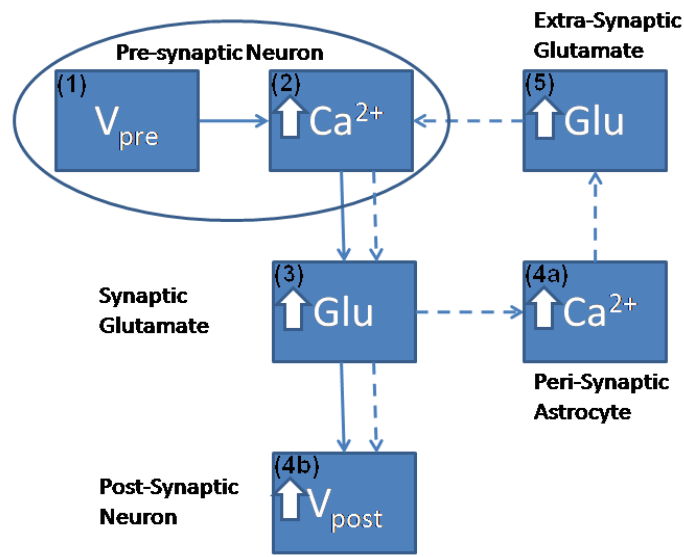

Figure 1. Information flow from pre-synaptic bouton to post-synaptic dendritic spine-head, as modulated by an astrocyte. Solid line shows the astrocyte-independent pathway, while, solid-line combined with dashed line shows the astrocyte-dependent pathway. (1) AP generated at pre-synaptic axon-hillock. (2) Elevated intracellular $[Ca^{2+}]$ in bouton. (3) Increased $[Ca^{2+}]$ leading to exocytosis of Glutamate into synaptic cleft. (4a) Synaptic glutamate causes an increase in astrocytic $[Ca^{2+}]$. (4b) Simultaneously synaptic glutamate can also bind with AMPAR causing an increase in post-synaptic membrane potential. (5) Increased astrocytic $[Ca^{2+}]$ leads to an elevated glutamate concentration in the extra-synaptic cleft, in a vesicle dependent manner. This extra-synaptic glutamate is free to bind with extra-synaptic mGluR on the pre-synaptic bouton surface. Glutamate bound to mGluR leads to an increase in $Ca^{2+}$ concentration via $IP_3$ dependent pathway. This transient enhancement of bouton $[Ca^{2+}]$ forms the basis of improved synaptic efficacy, through an astrocyte-dependent pathway.

## 2. The Model

In this section, we describe the details of the mathematical model, whose computational implementation will be presented in the section that immediately follows. In order to elucidate the major neurophysiological steps in our model we use the flow chart in Figure 1. The mathematical formulations have been described in the subsequent subsections.

2.1 Pre-synaptic Action Potential

Action potential (AP) is generated at the axon hillock of the pre-synaptic neuron. In the cortical neurons there may be eleven or more number of different ion channels (Lytton & Sejnowski, 1991). The key features of initiation dynamics of cortical neuron APs are (i) their rapid initiation and (ii) variable onset potential – are outside the range of behaviors described by the classical Hodgkin-Huxley (HH) theory (Naundorf et al., 2006). Still the HH paradigm has been used to generate pre-synaptic AP in computational models (Nadkarni & Jung, 2003; Volman et al., 2007). Since in this paper our focus is not on the detail of the pre-synaptic AP generation, for the sake of simplicity here we have followed the HH model for the pre-synaptic regular spikes and bursts generation.

$$C\frac{dV_{pre}}{dt} = I_{app} - g_K n^4(V_{pre} - V_K) - g_{Na} m^3 h(V_{pre} - V_{Na}) - g_L(V_{pre} - V_L)$$

$$\frac{dx}{dt} = \alpha_x(1-x) - \beta_x x$$

(1)

where $V_{pre}$ is the pre-synaptic membrane potential in millivolts, $I_{app}$ is the applied current density, $g_K$, $g_{Na}$ and $g_L$ are potassium, sodium and leak conductance respectively, $V_K$, $V_{Na}$ and $V_L$ are potassium, sodium and leak reversal potential respectively, and $x=m$ ($Na^+$ activation), $h$ ($Na^+$ inactivation) and $n$ ($K^+$ activation). The detail of the HH model can be found in (Hodgkin & Huxley, 1952). The values of the different parameters in equation (1) that have been used in this paper are furnished in the Table 2. $\alpha_x$ and $\beta_x$ for $x = m$, $h$ and $n$ are defined as

$$\alpha_n = \frac{0.01(-V_{pre} - 60)}{\exp(\frac{-V_{pre} - 60}{10}) - 1}, \alpha_m = \frac{0.1(-V_{pre} - 45)}{\exp(\frac{-V_{pre} - 45}{10}) - 1}, \alpha_h = 0.07\exp(\frac{-V_{pre} - 70}{20}),$$

$$\beta_n = 0.125\exp(\frac{-V_{pre} - 70}{80}), \beta_m = 4\exp(\frac{-V_{pre} - 70}{18}), \beta_h = \frac{1}{\exp(\frac{-V_{pre} - 40}{10}) + 1}$$

Table 2: Parameter values used in the HH model (all are from Hodgkin & Huxley, 1952)

| Symbol | Value |
|---|---|
| $g_K$ | 36 mS cm$^{-2}$ |
| $g_{Na}$ | 120 mS cm$^{-2}$ |
| $g_L$ | 0.3 mS cm$^{-2}$ |
| $V_K$ | −82 mV |
| $V_{Na}$ | 45 mV |
| $V_L$ | −59.4 mV |

2.2 Bouton Ca$^{2+}$ Dynamics

The train of AP that has been generated in the axon hillock of the pre-synaptic neuron, travels all the way down to the axon end feet without degradation and leads to an increase in cytosolic [Ca$^{2+}$]. The increase in intracellular [Ca$^{2+}$] can be attributed to two components:

i) [Ca$^{2+}$] due to AP, denoted as $c_{\text{fast}}$, and
ii) [Ca$^{2+}$] due to intracellular stores, $c_{\text{slow}}$.

Because of its rapid kinetics, [Ca$^{2+}$] due to AP is termed as $c_{\text{fast}}$. Similarly, [Ca$^{2+}$] due to intracellular stores is termed as $c_{\text{slow}}$. Total intracellular [Ca$^{2+}$] denoted as $c_i$ satisfies the following simple equation

$$c_i = c_{\text{fast}} + c_{\text{slow}} \Rightarrow \frac{dc_i}{dt} = \frac{dc_{\text{fast}}}{dt} + \frac{dc_{\text{slow}}}{dt} \qquad (2)$$

The sensitivity of rapid Ca$^{2+}$ kinetics over neurotransmitter release is well established (Schneggenburger & Neher, 2000; Bollman et al. 2000). In immature neurons, the necessary Ca$^{2+}$ flux for neurotransmitter release is primarily mediated by N-type Ca$^{2+}$ channels (Mazzanti & Haydon, 2003; Weber et al. 2010). Also, the contribution of P/Q- type channels is negligible as compared to N-type channels in immature cells (Ishikawa et al., 2006). Hence, in this article Ca$^{2+}$ influx through plasma membrane is modeled through N-type channels alone. Immature cells have been chosen following Perea & Araque (2007). The equation governing $c_{\text{fast}}$ consists of simple construction-destruction type formulism and is as follows (Keener & Sneyd, 1998)

$$\frac{dc_{\text{fast}}}{dt} = \underbrace{-\frac{I_{Ca} \cdot A_{btn}}{z_{Ca} F V_{btn}} + J_{PMleak}}_{\text{construction}} \underbrace{-\frac{I_{PMCa} \cdot A_{btn}}{z_{Ca} F V_{btn}}}_{\text{destruction}} \qquad (3)$$

Here, $I_{Ca}$ is the Ca$^{2+}$ current through N-type channel, $A_{btn}$ is the surface area of the bouton, $z_{Ca}$ is the Ca$^{2+}$ ion valence, $F$ is the Faraday's constant, $V_{btn}$ is the volume of the bouton. $I_{PMCa}$ represents the current due to electrogenic plasma-membrane Ca$^{2+}$ ATPase. This pump is known to extrude excess of Ca$^{2+}$ out of the cell and it has also been shown that it regulates excitatory synaptic transmission at CA3-CA1 pyramidal cell (CA3-CA1) synapse (Jensen et al., 2007). The formulation for this pump uses the standard Michaelis-Menton (MM) type formulism (Erler et al., 2004; Blackwell, 2005). $J_{PMleak}$ is the positive leak from extracellular space into bouton, which makes sure that MM pump does not decrease cytosolic Ca$^{2+}$ to 0 (Blackwell, 2005).

The Ca$^{2+}$ current through the N-type Ca$^{2+}$ channel is formulated using single protein level formulation, which is described in detail in (Erler et al. 2004)

$$I_{Ca} = \rho_{Ca} m_{Ca}^2 \underbrace{g_{Ca}(V_{pre}(t) - V_{Ca})}_{\text{Single open channel}}$$

Here, $\rho_{Ca}$ is the N-type channel protein density which determines the number of Ca$^{2+}$ channels on the membrane of the bouton ($\rho_{Ca}$ was determined computationally so that average neurotransmitter release probability lies in the range 0.2–0.3, when astrocyte is not stimulated, similar to the experiments of Perea & Araque (2007)), $g_{Ca}$ is the single N-type channel conductance, $V_{Ca}$ is the reversal potential of Ca$^{2+}$ ion determined by the Nernst equation (Keener & Sneyd, 1998),

$$V_{Ca} = \frac{RT}{z_{Ca}F} \ln\left(\frac{c_{ext}}{c_i^{rest}}\right) \quad (4)$$

where $R$ is the real gas constant, $T$ is the absolute temperature, $c_{ext}$ is the extracellular Ca$^{2+}$ concentration, $c_i^{rest}$ is the total intracellular [Ca$^{2+}$] at rest. It is assumed that a single N-type channel consists of two-gates. $m_{Ca}$ denotes the opening probability of a single gate. A single N-type channel is open only when both the gates are open. Hence, $m_{Ca}^2$ is the single channel open probability. The time dependence of the single channel open probability is governed by an HH-type formulation,

$$\frac{dm_{Ca}}{dt} = \frac{(m_{Ca}^\infty - m_{Ca})}{\tau_{m_{Ca}}}$$

where $m_{Ca}^\infty$ is the Boltzmann-function fitted by Ishikawa et al. (2005) to the whole cell current of an N-type channel, $m_{Ca}$ approaches its asymptotic value $m_{Ca}^\infty$ with a time constant $\tau_{m_{Ca}}$. The mathematical expression of other parameters used in equation (3) is as follows:

$$I_{PMCa} = v_{PMCa} \frac{c_i^2}{c_i^2 + K_{PMCa}^2}, \quad J_{PMleak} = v_{leak}(c_{ext} - c_i), \quad m_{Ca}^\infty = \frac{1}{1+\exp\left((V_{m_{Ca}} - V_m)/k_{m_{Ca}}\right)}.$$

Here, $v_{PMCa}$ is the maximum PMCa current density, determined through computer simulations, so that $c_i$ is maintained at its resting concentration. All other parameter values used for simulation are listed in Table 3.

Table 3: Parameters used for Bouton $Ca^{2+}$ dynamics

| Symbol | Description | Value | Reference |
| --- | --- | --- | --- |
| $F$ | Faraday's constant | 96487 C mole$^{-1}$ | |
| $R$ | Real gas constant | 8.314 J / K | |
| $T$ | Absolute Temperature | 293.15 K | Temperature in Perea & Araque (2007) |
| $z_{Ca}$ | Calcium valence | 2 | |
| $A_{btn}$ | Surface area of bouton | 1.24 μm$^2$ | Koester & Sakmann, 2000 |
| $V_{btn}$ | Volume of bouton | 0.13 μm$^3$ | Koester & Sakmann, 2000 |
| $\rho_{Ca}$ | N-type channel density | 3.2 μm$^{-2}$ | See text; Page No. 7 |
| $g_{Ca}$ | N-type channel conductance | 2.3 pS | Weber et al. 2010 |
| $V_{Ca}$ | Reversal potential of $Ca^{2+}$ ion | 125 mV | Calculated using equation (4) |
| $v_{PMCa}$ | Maximum PMCa current | 0.4 μA cm$^{-2}$ | See text; Page No. 8 |
| $K_{PMCa}$ | $Ca^{2+}$ concentration at which $v_{PMCa}$ is halved | 0.1 μM | Erler et al. 2004 |
| $v_{leak}$ | Maximum leak of $Ca^{2+}$ | 2.66 x 10$^{-6}$ ms$^{-1}$ | See Text; Page No. 7 |
| $c_i^{rest}$ | Resting Intracellular $Ca^{2+}$ concentration | 0.1 μM | Erler et al. 2004 |
| $c_{ext}$ | External $Ca^{2+}$ concentration | 2 mM | External [$Ca^{2+}$] in Perea & Araque (2007) |
| $V_{mCa}$ | Half-activation voltage of N-type $Ca^{2+}$ channel | -17 mV | Ishikawa et al. 2005 |
| $k_{mCa}$ | Slope factor of N-type channel activation | 8.4 mV | Ishikawa et al. 2005 |
| $\tau_{m_{Ca}}$ | Time constant of N-type channel | 10 ms | Ishikawa et al. 2005 |
| $c_1$ | Ratio of ER volume to volume of Bouton | 0.185 | Shuai & Jung, 2002 |
| $v_1$ | Maximum IP$_3$ receptor flux | 30 s$^{-1}$ | See text; Page No. 10 |
| $v_2$ | $Ca^{2+}$ leak rate constant | 0.055 s$^{-1}$ | See text; Page No. 10 |
| $v_3$ | SERCA maximal pump rate | 90 μM s$^{-1}$ | See text; Page No. 10 |
| $k_3$ | SERCA dissociation constant | 0.1 μM | Jafri & Keizer, 1995 |
| $d_1$ | IP$_3$ dissociation constant | 0.13 μM | Shuai & Jung, 2002 |
| $d_2$ | Inhibitory $Ca^{2+}$ dissociation constant | 1.049 μM | Shuai & Jung, 2002 |
| $d_3$ | IP$_3$ dissociation constant | 943.4 nM | Shuai & Jung, 2002 |
| $d_5$ | Activation $Ca^{2+}$ dissociation | 82.34 nM | Shuai & Jung, 2002 |

| | constant | | |
|---|---|---|---|
| $a_2$ | Inhibitory $Ca^{2+}$ binding constant | 0.2 μM s$^{-1}$ | Shuai & Jung, 2002 |
| $v_g$ | Maximum production rate of IP$_3$ | 0.062 μM s$^{-1}$ | Nadkarni & Jung, 2008 |
| $k_g$ | Glutamate concentration at which $v_g$ is halved | 0.78 nM | Nadkarni & Jung, 2008 |
| $\tau_p$ | IP$_3$ degradation constant | 0.14 s$^{-1}$ | Wang et al., 1995 |
| $p_0$ | Initial IP$_3$ concentration | 160 nM | Wang et al., 1995 |

The second component of bouton $Ca^{2+}$, $c_{slow}$, is the slower component. It is known to play a crucial role in STP (Emptage et al., 2001). The release of $Ca^{2+}$ from endoplasmic reticulum (ER) is mainly controlled by two types of receptors (or $Ca^{2+}$ channels) i) the inositol (1,4,5)-trisphosphate receptor (IP$_3$R) and ii) the ryanodine receptor (RyR) (Sneyd & Falcke, 2005). For the sake of simplicity, the flow is assumed to be through IP$_3$R alone. The IP$_3$ necessary for release of $Ca^{2+}$ from ER, is produced when glutamate (agonist) binds with mGluRs (receptor) and causes via G-protein link to phospholipase C (PLC), the cleavage of phosphotidylinositol (4,5)-bisphosphate (PIP$_2$) to produce IP$_3$ and diacylglycerol (DAG). We have used the conventional Li-Rinzel model (L-R model) (Li & Rinzel, 1994) to formulate this slower $Ca^{2+}$ signaling process.

There were a few modifications made to the L-R model. The L-R model assumes that, total intracellular concentration, $c_0$, is conserved and determines the ER $Ca^{2+}$ concentration, $c_{ER}$, using the following relation

$$c_{ER} = \frac{(c_0 - c_i)}{c_1}. \tag{5}$$

Such an assumption is not valid in the present model because of the presence of membrane fluxes, namely $I_{Ca}$ and $I_{PMCa}$. Also, in the L-R model intracellular IP$_3$ concentration, [IP$_3$], is used as a control parameter. To take care of these "inconveniences" two additional equations governing ER [$Ca^{2+}$] and [IP$_3$] have been incorporated in the L-R model. The [IP$_3$] production term was made glutamate dependent to study the effect of astrocytic $Ca^{2+}$ over $c_i$ (Nadkarni & Jung, 2007). The mathematical model governing the $c_{slow}$ dynamics is as follows

$$\frac{dc_{slow}}{dt} = -J_{chan} - J_{ERpump} - J_{ERleak},$$

$$\frac{dc_{ER}}{dt} = -\frac{1}{c_1}\frac{dc_{slow}}{dt},$$

$$\frac{dp}{dt} = v_g \frac{g_a^{0.3}}{k_g^{0.3} + g_a^{0.3}} - \tau_p (p - p_0),$$

$$\frac{dq}{dt} = \alpha_q (1-q) - \beta_q q.$$

(6)

Here $J_{chan}$ denotes Ca$^{2+}$ flux from ER to the intracellular space through IP$_3$R, $J_{ERpump}$ is the Ca$^{2+}$ flux pumped from the intracellular space into ER, $J_{ERleak}$ is the leak of Ca$^{2+}$ ions from ER to intracellular space, $c_{ER}$ is the ER Ca$^{2+}$ concentration, $c_1$ is the ratio of the volume of ER to the volume of bouton, $p$ is the intracellular IP$_3$ concentration, $g_a$ is the glutamate in the extra-synaptic cleft, $q$ is the fraction of activated IP$_3$R. The expressions for the fluxes are

$$J_{chan} = c_1 v_1 m_\infty^3 n_\infty^3 q^3 (c_i - c_{ER}),$$

$$J_{ERpump} = \frac{v_3 c_i^2}{k_3^2 + c_i^2},$$

$$J_{ERleak} = c_1 v_2 (c_i - c_{ER}),$$

with $m_\infty = \frac{p}{p+d_1}$, $n_\infty = \frac{c_i}{c_i+d_5}$, $\alpha_q = a_2 d_2 \frac{p+d_1}{p+d_3}$, $\beta_q = a_2 c_i$. Most of the values of $v_1$, $v_2$, $v_3$ mentioned in literature are for closed-cell dynamics which is not the case here. The values of $v_1$, $v_2$, $v_3$ were fixed through extensive simulation runs so that Ca$^{2+}$ homeostasis is maintained inside the cell and its organelles. Details of parameters are as listed in Table 3.

2.3 Glutamate release dynamics in bouton

It is now widely accepted that AP waveforms lead to a transient increase in intracellular [Ca$^{2+}$] and lead to neurotransmitter release (Bollman et al. 2000; Wang et al., 2009). However, the study of Ca$^{2+}$ sensor sensitivity becomes exceedingly challenging due to small size of nerve terminals (Wang et al., 2009). It is generally assumed that Ca$^{2+}$ concentration of at least 100 μM in the terminal is necessary for a "low-affinity" Ca$^{2+}$ sensor to activate (Neher, 1998; Nadkarni & Jung, 2008). But, recent studies performed at giant Calyx of Held terminal have revealed that intracellular Ca$^{2+}$ concentration of ~10 μM is sufficient for glutamate release (Schneggenburger & Neher, 2000; Bollman et al., 2000). The kinetic model governing the $Ca^{2+}$ binding to $Ca^{2+}$ sensor is given by the following equations (Bollman et al., 2000),

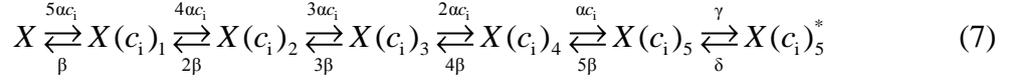

$$X \underset{\beta}{\overset{5\alpha c_i}{\rightleftarrows}} X(c_i)_1 \underset{2\beta}{\overset{4\alpha c_i}{\rightleftarrows}} X(c_i)_2 \underset{3\beta}{\overset{3\alpha c_i}{\rightleftarrows}} X(c_i)_3 \underset{4\beta}{\overset{2\alpha c_i}{\rightleftarrows}} X(c_i)_4 \underset{5\beta}{\overset{\alpha c_i}{\rightleftarrows}} X(c_i)_5 \underset{\delta}{\overset{\gamma}{\rightleftarrows}} X(c_i)_5^* \qquad (7)$$

Where, $\alpha$ and $\beta$ are the $Ca^{2+}$ association and dissociation rate constants respectively, $\gamma$ and $\delta$ are $Ca^{2+}$ independent isomerisation constants. $X$ is the $Ca^{2+}$ sensor (of a synaptic vesicle) with no $Ca^{2+}$ bound, $X(c_i)_1$ is $Ca^{2+}$ sensor with one $Ca^{2+}$ bound, likewise, $X(c_i)_5$ is $Ca^{2+}$ sensor with five $Ca^{2+}$ bound; $X(c_i)_5^*$ is the isomer of $X(c_i)_5$ which is ready for glutamate release. Hippocampal synapses are known as low-fidelity synapses (Nadkarni & Jung, 2008). We have assumed an active zone consisting of two-docked synaptic vesicles (Danbolt, 2001; Nikonenko & Skibo, 2006). Since there are few synaptic vesicles the number of vesicles with 5 $Ca^{2+}$ ions bound cannot be determined by the average of vesicle pool. Therefore, fraction of docked vesicles ready to be released $f_r$, has been determined using dynamic Monte-Carlo simulation (Fall et al., 2002) of kinetic equation (7) and depends on $X(c_i)_5^*$ state.

Apart from evoked release of docked vesicles, spontaneous release of vesicles can also occur when pre-synaptic membrane is not depolarized. The rate of spontaneous release depends upon pre-synaptic $Ca^{2+}$ concentration (Bollman et al., 2000; Emptage et al., 2001; Schneggenburger & Neher, 2000). The number of vesicles ready to be released spontaneously, $p_r$, is assumed to be a Poisson process with the following rate,

$$\lambda(c_i) = a_3 \left(1 + \exp\left(\frac{a_1 - c_i}{a_2}\right)\right)^{-1}. \qquad (8)$$

The formulation for the rate of spontaneous release is from Nadkarni & Jung (2008). We have to modify the parameter values (see equation (8)) because as per their choice of values and system setup, the frequency of spontaneously released vesicles was as high as 19 per sec (we have determined this through simulation runs of over 10000 trials). However, the experimentally determined frequency of spontaneous vesicle release in presence of an astrocyte is in between $1 - 3$ per sec (Kang et al., 1998). Thus, we determined the values of $a_1$, $a_2$ and $a_3$ by simulation so that the frequency of spontaneous vesicle release is between $1 - 3$ Hz. The vesicle fusion and recycling process is governed by the Tsodyks & Markram Model (TMM) (Tsodyks &

Markram, 1997). A slight modification has been made to the TMM to make the vesicle fusion process $p_r$ dependent. The modified TMM is as follows

$$\frac{dR}{dt} = \frac{I}{\tau_{rec}} - f_r \cdot R,$$
$$\frac{dE}{dt} = -\frac{E}{\tau_{inact}} + f_r \cdot R, \qquad (9)$$
$$I = 1 - R - E,$$

where $R$ is the fraction of releasable vesicles inside bouton, $E$ is the fraction of effective vesicles in the synaptic cleft and $I$ is the fraction of inactive vesicles undergoing recycling process, $f_r$ has the values (0, 0.5, 1) corresponding to the number of vesicles ready to be released (0, 1, 2), which is determined by the stochastic simulation of kinetic model in equation (7) or generating a Poisson random variable with the rate given by the equation (8). $\tau_{inact}$ and $\tau_{rec}$ are the time constants of vesicle inactivation and recovery respectively. Once a vesicle is released whether evoked or spontaneous the vesicle release process remains inactivated for a period of 6.34 ms (Dobrunz et al., 1997). The parametric values used for simulation are listed in Table 4.

2.4 Glutamate dynamics in synaptic cleft

Various types of glutamate receptors have been detected pre-synaptically, extra-synaptically, as well as on glial cells (Danbolt, 2001). Suggesting that, to study transmission of glutamatergic signals, it is essential to study, how glutamate diffuses (Danbolt, 2001). However, using Monte Carlo simulation of a central glutamatergic synapse, in particular a CA3–CA1 synapse, Franks et al., (2002) showed that glutamatergic signaling is spatially independent at these synapses. The capacity of the bouton vesicle containing glutamate has been assumed to be 60 mM (Danbolt, 2001). Since, E gives the effective fraction of vesicles in the cleft the estimated glutamate concentration in the cleft can be represented mathematically as

$$\frac{dg}{dt} = n_v \cdot g_v \cdot E - g_c \cdot g. \qquad (10)$$

Here $g$ is the glutamate concentration in the synaptic cleft, $n_v$ is the number of docked vesicle, $g_v$ is the vesicular glutamate concentration and $g_c$ is the rate of glutamate clearance i.e. re-uptake by neuron or astrocyte (Destexhe et al., 1998). Using this simple dynamics, we could achieve the estimated range of glutamate concentration 0.24 - 11 mM in cleft (Danbolt, 2001; Franks et al., 2002) and the time course of glutamate in the cleft is 2 ms (Clements, 1996; Franks et al., 2002). Although similar equation can be used to model glutamate dynamics at other synapses, however, one might have to use different constant values. Thus, the present formulation can be considered specific to a CA3–CA1 synapse.

Table 4: Parameters used for Glutamate dynamics in bouton and cleft

| Symbol | Description | Value | Reference |
|---|---|---|---|
| $\alpha$ | $Ca^{2+}$ association rate constant | 0.3 μM ms$^{-1}$ | Bollman et al. 2000 |
| $\beta$ | $Ca^{2+}$ dissociation rate constant | 3 ms$^{-1}$ | Bollman et al. 2000 |
| $\gamma$ | Isomerization rate constant (forward) | 30 ms$^{-1}$ | Bollman et al. 2000 |
| $\delta$ | Isomerization rate constant (backward) | 8 ms$^{-1}$ | Bollman et al. 2000 |
| $\tau_{rec}$ | Vesicle recovery time constant | 800 ms | Tsodyks & Markram, 1997 |
| $\tau_{inac}$ | Vesicle inactivation time constant | 3 ms | Tsodyks & Markram, 1997 |
| $a_1$ | $Ca^{2+}$ concentration at which $\lambda$ is halved | 50 μM | See text; Page No. 11 |
| $a_2$ | Slope factor of spontaneous release rate $\lambda$ | 5 μM | See text; Page No. 11 |
| $a_3$ | Maximum spontaneous release rate | 0.85 ms$^{-1}$ | See text; Page No. 11 |
| $n_v$ | Number of docked vesicle | 2 | Nikonenko & Skibo, 2006 |
| $g_v$ | Glutamate concentration in single vesicle | 60 mM | Montana et al., 2006 |
| $g_c$ | Glutamate clearance rate constant | 10 ms$^{-1}$ | Destexhe et al., 1998 |

## 2.5 Astrocyte $Ca^{2+}$ dynamics

Porter & McCarthy (1996) showed that glutamate released from the Schaffer collaterals (SC) leads to an increase in astrocytic $Ca^{2+}$ via an mGluR pathway. Recently, De Pitta et al. (2009) proposed a G-ChI model for astrocytic $Ca^{2+}$ oscillations mediated by mGluR pathway while treating glutamate concentration in the synaptic cleft as a parameter. They called it G-ChI referring to the dependent variables and the glutamate concentration parameter used in their model (in their model G represented glutamate concentration in the synaptic cleft, C represented astrocytic $[Ca^{2+}]$, h represented the gating variable of IP$_3$R and I represented the astrocytic $[IP_3]$). We have used the G-ChI model for astrocyte $Ca^{2+}$ dynamics with an exception that '$g$' is a dynamic variable given by equation (10). The G-ChI model uses the conventional L-R model for astrocytic $Ca^{2+}$ concentration $c_a$ with some specific terms for intracellular IP$_3$ concentration $p_a$. It incorporates PLC**β** and PLC**δ** (are isoenzymes of the family of phosphoinositide specific PLC) dependent IP$_3$ production. It also incorporates inositol polyphosphate 5-phosphatase (IP-5P) and IP$_3$ 3-kinase (IP$_3$-3K) dependent IP$_3$ degradation (for a systematic derivation regarding the expressions, shown in equation (13), incorporating these effects see De Pitta et al., 2009). It is a very detailed model based on astrocyte specific experiments (Hofer et al., 2002; Suzuki et al., 2004), model which exhibits IP$_3$ oscillations apart from Ca$^{2+}$ oscillations. However, the exact significance of IP$_3$ oscillations is yet unknown (De Pitta et al., 2009). The G-Chi model is a closed-cell model (Keener & Sneyd, 2009) i.e. without membrane fluxes. In such models $c_a$ primarily depends upon two parameters, i) flux from ER into cytosol and ii) the maximal pumping capacity of the Sarco-Endoplasmic Reticulum ATPase (SERCA) pump. It is known that IP$_3$Rs are found in clusters in astrocytes (Holtzclaw et al., 2002). However, the size of the cluster $N_{IP3}$ is not known. We have assumed it to be 20 (Shuai & Jung, 2002). We make use of the stochastic L-R model (Shuai & Jung, 2002). The model can be represented as follows

$$\frac{dc_a}{dt} = \left(r_{c_a} m_\infty^3 n_\infty^3 h_a^3\right)\left(c_0 - (1+c_{1,a})c_a\right) - v_{ER}\frac{c_a^2}{c_a^2 + K_{ER}^2} + r_L\left(c_0 - (1+c_{1,a})c_a\right), \quad (11)$$

$$\frac{dp_a}{dt} = v_\beta \cdot \text{Hill}\left(g^{0.7}, K_R\left(1+\frac{K_p}{K_R}\text{Hill}(C, K_\pi)\right)\right) + \frac{v_\delta}{1+\frac{p_a}{k_\delta}}\text{Hill}\left(c_a^2, K_{PLC\delta}\right) \quad (12)$$

$$- v_{3K}\text{Hill}\left(c_a^4, K_D\right)\text{Hill}\left(p_a, K_3\right) - r_{5p_a} p_a$$

$$\frac{dh_a}{dt} = \alpha_{h_a}(1-h_a) - \beta_{h_a} h_a + G_h(t) \quad (13)$$

Here the first term on the right hand side of equation (11) represents the Ca$^{2+}$ flux flowing out from ER to the intracellular space, the second term represents the rate at which Ca$^{2+}$ is removed from the intracellular space by SERCA pump and the last term represents the leak of Ca$^{2+}$ from ER into the intracellular space. Clearly these terms are very analogous to the terms involved in production of $c_{slow}$ in equation (6). But with a major difference, which was mentioned earlier as well, that this model is based on closed-cell assumption. Under such an assumption, an expression like equation (5) holds true and can be represented in terms of the astrocyte cell parameters as

$$c_{ER,a} = \frac{(c_0 - c_a)}{c_{1,a}} \Rightarrow c_{ER,a} c_{1,a} = c_0 - c_a. \quad (14)$$

Equation (14) gives us the advantage to represent astrocytic Ca$^{2+}$ flux terms completely in terms of cell parameters (compare equation (11) with equation (6) where a separate differential equation for $dc_{ER}/dt$ is present). $r_{c_a}$ is the maximal rate of Ca$^{2+}$ flux from IP$_3$R cluster, $m_\infty^3 n_\infty^3 h_a^3$ together represent the opening probability of IP$_3$R cluster. $v_{ER}$ is the maximal rate of Ca$^{2+}$ uptake into ER, $K_{ER}$ is the affinity of SERCA pump for intracellular Ca$^{2+}$. $r_L$ is the maximal rate of Ca$^{2+}$ leak from ER. The first two terms on the right hand side of equation (12) incorporate agonist-dependent and agonist-independent production of IP$_3$ and the last two terms incorporate IP$_3$ degradation by IP$_3$-3K and IP-5P respectively. In equation (13), $\alpha_{h_a}$ is the rate at which $h_a$, $\beta_{h_a}$ is the rate at which $h_a$ closes and G$_h$(t) is zero mean, uncorrelated, Gaussian white-noise term with co-variance function (Shuai & Jung, 2002),

$$\langle G_h(t) G_h(t') \rangle = \frac{\alpha_{h_a}(1-h_a) + \beta_{h_a} h_a}{N_{IP_3}} \delta(t-t')$$

Here, $\delta(t)$ is the Dirac-delta function, $t$ and $t'$ are distinct time and $\frac{\alpha_{h_a}(1-h_a) + \beta_{h_a} h_a}{N_{IP_3}}$ is the spectral density (Coffey et al. 2005). The present model can be classified into three categories i) amplitude modulated (AM), ii) frequency modulated (FM), and iii) amplitude and frequency modulated (AFM) modulated (De Pitta et al., 2009). We have used AFM-encoded astrocytic Ca$^{2+}$ oscillations as coupling of IP$_3$ metabolism with calcium-induced calcium release (CICR) does not allow pure AM encoding (De Pitta et al., 2009). The mathematical expression of other parameters used in equations (11) and (13) are

$$m_{\infty,a} = \text{Hill}(p_a, d_1), \ n_{\infty,a} = \text{Hill}(c_a, d_5), \ \text{Hill}(x^n, K) = \frac{x^n}{x^n + K^n},$$

$$\alpha_{h_a} = a_2 d_2 \frac{p_a + d_1}{p_a + d_3}, \beta_{h_a} = a_2 c_a.$$

$\text{Hill}(x^n, K)$ is the generic Hill function (De Pitta et al., 2009). Typically, Hill function is used for reactions whose intermediate steps are unknown (or not considered) but cooperative behavior is suspected in the reaction (Keener & Sneyd, 1998). Mathematically, it can be said that Hill function is used for reactions whose reaction velocity curve is not hyperbolic (Keener & Sneyd, 1998). Parametric value of all the constants is as listed in Table 5.

Table 5: Parameters used for astrocyte Ca$^{2+}$ dynamics

| Symbol | Description | Value | Reference |
|--------|-------------|-------|-----------|
| $r_{c_a}$ | Maximal IP$_3$R flux | 6 s$^{-1}$ | De Pitta et al. 2009 |
| $r_L$ | Maximal rate of Ca$^{2+}$ leak from ER | 0.11 s$^{-1}$ | De Pitta et al. 2009 |
| $c_0$ | Total cell free Ca$^{2+}$ concentration | 2 μM | De Pitta et al. 2009 |
| $c_{1,a}$ | Ratio of ER volume to cytosol volume | 0.185 | De Pitta et al. 2009 |
| $v_{ER}$ | Maximal rate of SERCA uptake | 0.9 μM s$^{-1}$ | De Pitta et al. 2009 |
| $K_{ER}$ | SERCA Ca$^{2+}$ affinity | 0.1 μM | De Pitta et al. 2009 |
| $d_1$ | IP$_3$ dissociation constant | 0.13 μM | De Pitta et al. 2009 |
| $d_2$ | $Ca^{2+}$ inactivation dissociation constant | 1.049 μM | De Pitta et al. 2009 |
| $d_3$ | $IP_3$ dissociation constant | 0.9434 μM | De Pitta et al. 2009 |
| $d_5$ | $Ca^{2+}$ activation dissociation constant | 0.08234 μM | De Pitta et al. 2009 |
| $a_2$ | $IP_3R$ binding rate for $Ca^{2+}$ Inhibition | 2 $s^{-1}$ | De Pitta et al. 2009 |
| $N$ | Number of $IP_3R$ in a cluster | 20 | Shuai & Jung, 2002 |
| | Glutamate-dependent $IP_3$ production | | |
| $v_\beta$ | Maximal rate of $IP_3$ production by PLCβ | 0.5 μM $s^{-1}$ | De Pitta et al. 2009 |
| $K_R$ | Glutamate affinity of the receptor | 1.3 μM | De Pitta et al. 2009 |
| $K_p$ | $Ca^{2+}$/PKC-dependent inhibition factor | 10 μM | De Pitta et al. 2009 |
| $K_\pi$ | $Ca^{2+}$ affinity of PKC | 0.6 μM | De Pitta et al. 2009 |
| | Glutamate-independent $IP_3$ production | | |
| $v_\delta$ | Maximal rate of $IP_3$ production by PLCδ | 0.05 μM $s^{-1}$ | De Pitta et al. 2009 |
| $K_{PLC\delta}$ | $Ca^{2+}$ affinity of PLCδ | 0.1 μM | De Pitta et al. 2009 |
| $k_\delta$ | Inhibition constant of PLCδ activity | 1.5 μM | De Pitta et al. 2009 |
| | $IP_3$ degradation | | |
| $r_{5pa}$ | Maximal rate of degradation by IP-5P | 0.05 $s^{-1}$ | De Pitta et al. 2009 |
| $v_{3K}$ | Maximal rate of degradation by $IP_3$-3K | 2 μM $s^{-1}$ | De Pitta et al. 2009 |
| $K_D$ | $Ca^{2+}$ affinity of $IP_3$-3K | 0.7 μM | De Pitta et al. 2009 |
| $K_3$ | $IP_3$ affinity of $IP_3$-3K | 1 μM | De Pitta et al. 2009 |

2.6 Gliotransmitter release dynamics in astrocyte

There is enough evidence that astrocytes actually release gliotransmitters in a $Ca^{2+}$ dependent manner (Bezzi et al. 2004; Montana et al. 2006; Bowser & Khakh, 2007; Marchaland et al. 2008; Fellin, 2009). There is again considerable evidence that the released gliotransmitters modulate synaptic plasticity via extra-synaptic NMDAR (Parpura et al. 1994; Parpura & Haydon, 2000; Carmignoto & Fellin, 2006; Bergersen & Gundersen, 2009) and extra-synaptic mGluR (Fiacco & McCarthy, 2004; Perea & Araque, 2007). But, the exact mechanism by which astrocytes release gliotransmitters is yet to be determined (Wenker, 2010). However, it is widely agreed upon that astrocytes release gliotransmitters in a vesicular manner similar to neurons (Bezzi et al. 2004; Montana et al., 2006; Verkhratsky & Butt, 2007; Marchaland et al. 2008) as they possess the necessary exocytotic secretory machinery (Parpura & Zorec, 2010). In 2000, Parpura & Haydon determined $Ca^{2+}$ dependency of glutamate release from hippocampal astrocytes and found that the Hill coefficient for glutamate release from astrocytes was 2.1–2.7 suggesting at least two $Ca^{2+}$ ions are must for a possible gliotransmitter release. Recently the probability of vesicular fusion in response to a mechanical stimulation and the size of readily releasable pool of SLMVs in astrocytes have been determined by Malarkey & Parpura (2011). Based on the observation of Parpura & Haydon (2000) in this manuscript we have assumed that binding of three $Ca^{2+}$ ions leads to a gliotransmitter release. The model governing the gliotransmitter release site activation is based on Bertram et al. (1996). Our gliotransmitter release model assumes that three $Ca^{2+}$ ions must bind with three independent gates or sites ($S_1$ – $S_3$) for a possible gliotransmitter release.

$$c_a + C_j \underset{k_j^-}{\overset{k_j^+}{\rightleftarrows}} O_j, \qquad j = 1, 2, 3,$$

where $C_j$ and $O_j$ are the closing and opening probability of gate $S_j$ respectively, $k_j^+$ and $k_j^-$ are the opening and closing rates of the gate $S_j$ respectively. The temporal evolution of the open gate $O_j$ can be expressed as

$$\frac{dO_j}{dt} = k_j^+ \cdot c_a - \left(k_j^+ \cdot c_a + k_j^-\right) \cdot O_j. \tag{15}$$

As the three sites are physically independent, the fraction of SLMVs ready to be released can be given as the product of the opening probabilities of the three sites

$$f_r^a = O_1 \cdot O_2 \cdot O_3. \tag{16}$$

The dissociation constants of gates $S_1$ – $S_3$ are 108 nM, 400 nM, and 800 nM. The time constants for gate closure ($1/k_j^-$) are 2.5 s, 1s, and 100 ms. The dissociation constants and time constants for $S_1$ and $S_2$ are same as in Bertram et al (1996). While, the dissociation constant and time constant for gate $S_3$ was fixed through computer simulations to fit the experimentally determined probability of fusogenic (fraction of readily releasable SLMVs in response to a mechanical stimulation) SLMVs found recently by Malarkey & Parpura (2011). Once an SLMV is ready to be released its fusion and recycling process was modeled using TMM. The governing model is as follows

$$\begin{aligned}
\frac{dR_a}{dt} &= \frac{I_a}{\tau_{rec}^a} - \Theta\left(c_a - c_a^{thresh}\right) \cdot f_r^a \cdot R_a, \\
\frac{dE_a}{dt} &= -\frac{E_a}{\tau_{inact}^a} + \Theta\left(c_a - c_a^{thresh}\right) \cdot f_r^a \cdot R_a, \\
I_a &= 1 - R_a - E_a.
\end{aligned} \tag{17}$$

Here, $R_a$ is the fraction of readily releasable SLMVs inside the astrocyte, $E_a$ is the fraction of effective SLMVs in the extra-synaptic cleft and $I_a$ is the fraction of inactive SLMVs undergoing endocytosis or re-acidification process. $\Theta$ is the Heaviside function and $c_a^{thresh}$ is the threshold of astrocytic [Ca$^{2+}$] necessary for release site activation (Parpura & Haydon, 2000). $\tau_{inact}^a$ and $\tau_{rec}^a$ are the time constants of inactivation and recovery of SLMVs respectively.

2.7 Glutamate dynamics in extra-synaptic cleft

The glutamate concentration in the extra-synaptic cleft $g_a$, has been modeled in a similar way to equation (10). This glutamate acts on extra-synaptically located mGluRs of the pre-synaptic bouton. It is used as an input in the IP$_3$ production term of equation (6). The SLMVs of the astrocytes are not as tightly packed as of the neurons (Bezzi et al., 2004). Thus, it is assumed that each SLMV contains 20 mM of glutamate (Montana et al., 2006). The mathematical equation governing glutamate dynamics in the extra-synaptic cleft are as follows

$$\frac{dg_a}{dt} = n_a^v \cdot g_a^v \cdot E_a - g_a^c \cdot g_a, \tag{18}$$

where $g_a$ is the glutamate concentration in the extra-synaptic cleft, $n_a^v$ represents the readily releasable pool of SLMVs, $g_a^v$ is the glutamate concentration within each SLMV, $g_a^c$ is the clearance rate of glutamate from the cleft due to diffusion and/or re-uptake by astrocytes.

Table 6: Parameters used for Glutamate dynamics in astrocyte and extra-synaptic cleft

| Symbol | Description | Value | Reference |
|---|---|---|---|
| $k_1^+$ | Ca$^{2+}$ association rate for S$_1$ | 3.75 x 10$^{-3}$ μM$^{-1}$ ms$^{-1}$ | Bertram et al. 1996 |
| $k_1^-$ | Ca$^{2+}$ dissociation rate for S$_1$ | 4 x 10$^{-4}$ ms$^{-1}$ | Bertram et al. 1996 |
| $k_2^+$ | Ca$^{2+}$ association rate for S$_2$ | 2.5 x 10$^{-3}$ μM$^{-1}$ ms$^{-1}$ | Bertram et al. 1996 |
| $k_2^-$ | Ca$^{2+}$ dissociation rate for S$_2$ | 1 x 10$^{-3}$ ms$^{-1}$ | Bertram et al. 1996 |
| $k_3^+$ | Ca$^{2+}$ association rate for S$_3$ | 1.25 x 10$^{-2}$ μM$^{-1}$ ms$^{-1}$ | See text, page no. 17 |
| $k_3^-$ | Ca$^{2+}$ dissociation rate for S$_3$ | 10 x 10$^{-3}$ ms$^{-1}$ | See text, page no. 17 |
| $\tau_{rec}^a$ | Vesicle recovery time constant | 800 ms | Tsodyks & Markram, 1997 |
| $\tau_{inac}^a$ | Vesicle inactivation time constant | 3 ms | Tsodyks & Markram, 1997 |
| $c_a^{thresh}$ | Astrocyte response threshold | 196.69 nM | Parpura & Haydon, 2000 |
| $n_a^v$ | SLMV ready to be released | 12 | Malarkey & Parpura, 2011 |

| $g_a^v$ | Glutamate concentration in one SLMV | 20 mM | Montana et al. 2006 |
| $g_a^c$ | Glutamate clearance rate from the extra-synaptic cleft | 10 ms$^{-1}$ | Destexhe et al. 1998 |

2.8 Dendritic spine-head dynamics

The dendritic spine-head is assumed to be of mushroom type. Its volume is taken to be 0.9048 μm$^3$ (assuming a spherical spine-head of radius 0.6 μm (Dumitriu et al., 2010)). The specific capacitance and specific resistance of the spine-head is assumed to be 1 μF / cm$^2$ and 10000 Ω cm$^2$, respectively. Given the dimension of the spine we can calculate its actual resistance as

$$R_m = \frac{R_M}{A_{spine}}, \qquad (19)$$

where $R_m$ is the actual resistance of the spine, $R_M$ is the specific resistance of the spine and $A_{spine}$ is the area of spine-head. NMDAR (N-methyl D-aspartate receptor) and AMPAR (α-amino-3-hydroxy-5-methyl-4-isoxazolepropionic acid receptor) are co-localized at most of the glutamatergic synapses, most of which are found at dendritic spines (Franks et al., 2002). Chen & Diamond (2002) showed that the post-synaptic NMDAR receives less glutamate during evoked synaptic response, suggesting that most of the post-synaptic current is contributed by AMPAR, under such conditions. Also, NMDAR is known to play a crucial role in longer forms of synaptic plasticity, Long-term Potentiation (LTP) and Long-term Depression (LTD) (Bliss & Collingridge, 1993; Malenka & Bear, 2004). Hence, in our model of short-term potentiation the post-synaptic density comprises of AMPAR alone. The post-synaptic potential change has been modeled using a passive membrane mechanism (Tsodyks & Markram, 1997)

$$\tau_{post} \frac{dV_{post}}{dt} = -(V_{post} - V_{post}^{rest}) - R_m \cdot I_{AMPA}, \qquad (20)$$

where $\tau_{post}$ is the post-synaptic membrane time constant, $V_{post}^{rest}$ is the post-synaptic resting membrane potential, $I_{AMPA}$ is the AMPAR current and is given by the following expression (Destexhe et al., 1998)

$$I_{AMPA} = g_{AMPA} m_{AMPA} \left( V_{post} - V_{AMPA} \right),$$

where $g_{AMPA}$ is the conductance of the AMPAR channel, $V_{AMPA}$ is the reversal potential of the AMPAR and $m_{AMPA}$ is the gating variable of AMPAR. Although there exists a more comprehensive 6-state markov model for AMPAR gating (Destexhe et al., 1998), in our model we have used a simple 2-state model for AMPAR gating. This two state model has been used keeping in mind it is computationally less expensive, while retaining most of the important qualitative properties (Destexhe et al., 1998). Also, it is known that detailed AMPAR mechanisms like desensitization do not play a role in STP (Zucker & Regehr, 2002). AMPAR gating is governed by the following HH-type formulism (Destexhe et al., 1998)

$$\frac{dm_{AMPA}}{dt} = \alpha_{AMPA} g (1 - m_{AMPA}) - \beta_{AMPA} m_{AMPA}. \tag{21}$$

Here, $\alpha_{AMPA}$ is the opening rate of the receptor, $\beta_{AMPA}$ is the closing rate of the receptor and $g$ is the glutamate concentration in the cleft given by equation (10). The parameter values are as listed in Table 7.

Table 7: List of parameters used for post-synaptic potential generation

| Symbol | Description | Value | Reference |
|---|---|---|---|
| $R_m$ | Actual resistance of the spine-head | $0.79 \times 10^5$ M$\Omega$ | Calculated using equation (19) |
| $V_{post}^{rest}$ | Post-synaptic resting membrane potential | -70 mV | |
| $\tau_{post}$ | Post-synaptic membrane time constant | 50 ms | Tsodyks & Markram, 1997 |
| $g_{AMPA}$ | AMPAR conductance | 0.35 nS | Destexhe et al. 1998 |
| $V_{AMPA}$ | AMPAR reversal potential | 0 mV | Destexhe et al. 1998 |
| $\alpha_{AMPA}$ | AMPAR forward rate constant | 1.1 μM s$^{-1}$ | Destexhe et al. 1998 |
| $\beta_{AMPA}$ | AMPAR backward rate constant | 190 s$^{-1}$ | Destexhe et al. 1998 |

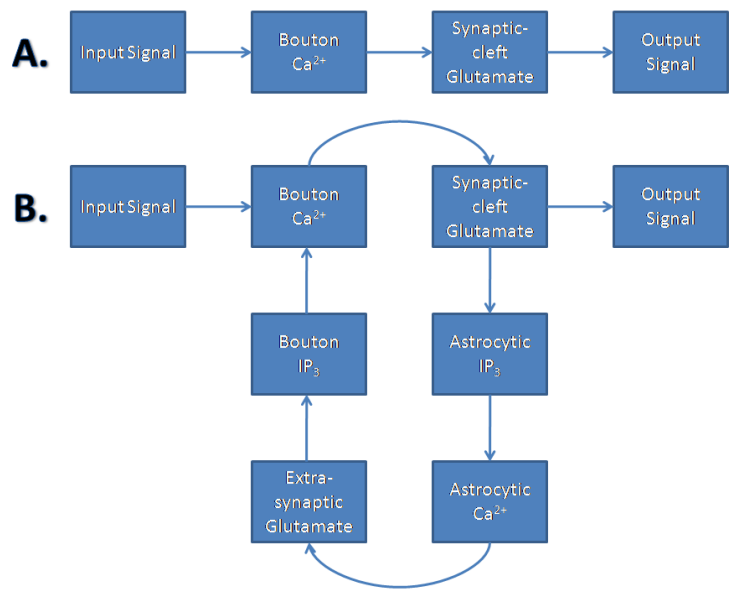

Figure 2. The two types of information processing simulated in this paper. (A) Astrocyte-independent information processing. (B) Astrocyte-dependent information processing. The input signal is being amplified by astrocyte-dependent feed-forward and feed-back pathways making up a loop.

2.9 Numerical Implementation

All the computations have been performed using MATLAB. The model equations were discretized with a temporal precision of Δt = 0.05 ms. The canonical explicit Euler method was used to solve the system of twenty-two ordinary differential equations governing TpS. For the numerical simulation of the noise term, in equation (13), we have used Box-Muller Algorithm (Fox, 1997) to generate noise-term at each time-step (Δ*t*). All simulations were performed on a Dell precision 3500 workstation with Intel Xeon processor with 2.8 GHz processing speed and with 12 GB RAM. The time taken for model time of 1s (stimulation rate 5 Hz) is approximately 8.5 sec. The MATLAB script written for the simulation of the model can be requested by email to any of the authors.

3. Simulation results

How post-synaptic current is being generated with and without the participation of astrocytic $Ca^{2+}$ have been shown in this section with extensive numerical simulations of the model equations presented in the previous section. In the latter case how the output signal is being amplified through a processing loop, consisting of feed-forward and feed-back paths, with the help of astrocytic $Ca^{2+}$ signaling, has been shown in Figure 2B. Here, we have tried to answer the question, "Does astrocyte play an active role in modulation of synaptic plasticity?" In order to study the difference in both types of processing (see Figure 2), first we present the results associated with astrocyte-independent processing followed by astrocyte-dependent processing.

3.1 Astrocyte-independent Information Processing

In this subsection we simulate the processing elaborated in Figure 2A. We present results of implementation of the models described in subsections 2.1, 2.2, 2.4 and 2.8 (Figures 3A, 3B, 3C, and 3D respectively).

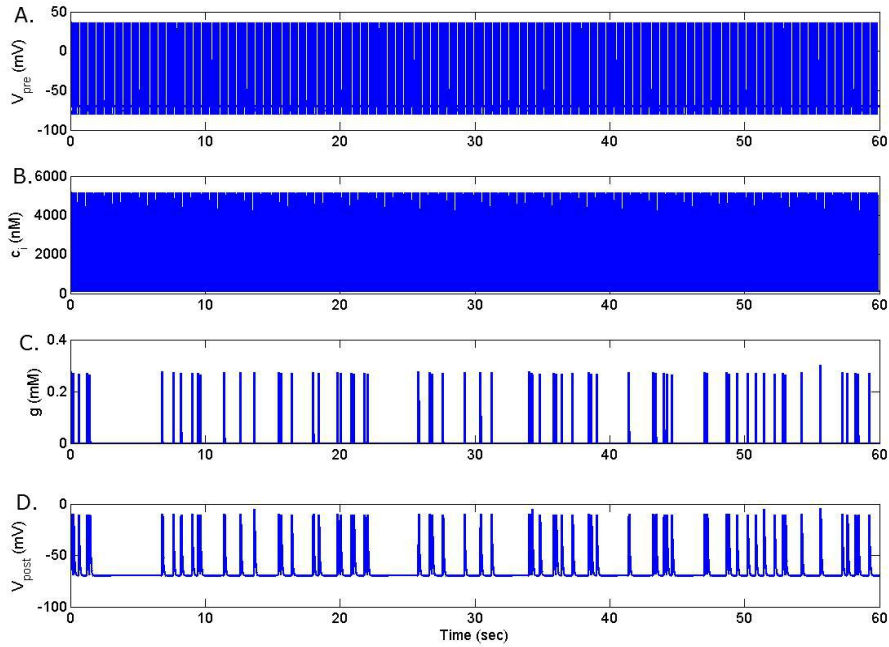

Figure 3. The major variables involved in astrocyte-independent information processing. A. $V_{pre}$ (mV), 5 Hz input signal generated using HH model, in response to a stimulus of 10 μA per cm$^2$ of frequency 5 Hz and duration 10 ms. B. Ca$^{2+}$ (nM), fast Ca$^{2+}$ oscillations in response to the 5 Hz input signal. C. Synaptic glutamate (mM), elevated glutamate concentration in the synaptic cleft due to exocytosis of glutamate filled synaptic vesicles from bouton. D. Excitatory post-synaptic potential (EPSP) (mV), potential change in the membrane of the post-synaptic spine mediated through AMPAR channels.

We used the model described in equation (1) to generate input signal or pre-synaptic membrane potential. This input signal forms the basis of signal transduction and we made sure that the tripartite synapse is at rest in its absence. In response to this input signal, the N-type Ca$^{2+}$ channels open and bouton Ca$^{2+}$ starts undergoing very fast oscillations (see Figure 3B). Note that, here, there is no astrocyte present and hence there is no contribution of [Ca$^{2+}$] from intracellular stores. We adjusted the number of Ca$^{2+}$ channels on the surface of the bouton (by adjusting $\rho_{Ca}$) so that the amplitude of Ca$^{2+}$ oscillation is 5 μM i.e., exactly half of the affinity of Ca$^{2+}$ sensor (β/α, where β and α are given in Table 4). Doing this we could attain average neurotransmitter release probability, in the range 0.2–0.3 (see Figure 5), observed experimentally in absence of astrocyte (Perea & Araque, 2007). Increased bouton [Ca$^{2+}$] instigates the process of exocytosis and vesicles release their content (glutamate) in the synaptic cleft (see Figure 3C). When glutamate concentration rises in the cleft, it binds with post-synaptic AMPAR, which causes this ligand-gated channel to open. Once opened, AMPAR causes a change in the post-synaptic potential (see Figure 3D) since this deflection is positive it has been termed as EPSP. As described in the previous section, we also keep track of the vesicle recycling process, see equation (9), which is shown in Figure 4.

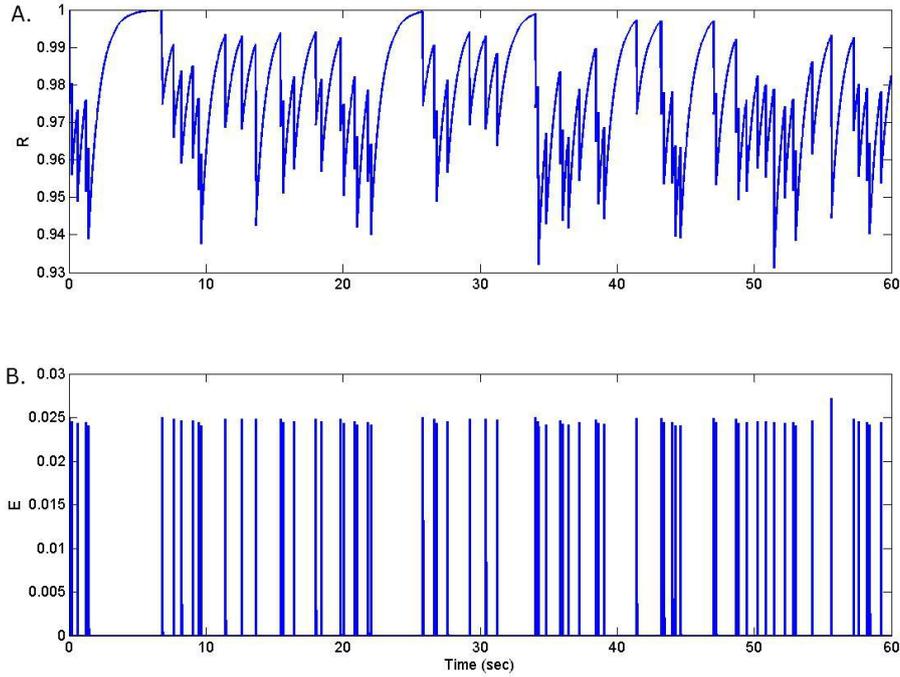

Figure 4. Fraction of releasable and effective vesicles, in astrocyte-independent information processing, during an input signal of 5 Hz (see Figure 3A). (A) The fraction of releasable vesicles i.e., ready to be fused, inside the bouton. (B) The fraction of effective vesicles i.e., fraction of vesicles left in the synaptic cleft.

In Figure 4 we show the underlying process of vesicle release. In the absence of astrocyte, it can be observed that nearly 90% of the vesicles are available for release for most of the time (see Figure 4A). In Figure 4B, we observe that the fraction of effective vesicles is not as dense as the input signal (see Figure 3A) implying low probability of vesicle release. In fact, the probability of vesicle release was nearly 0.25 i.e., every fourth input signal is able to release a synaptic vesicle. We next show Pr i.e., neurotransmitter release probability in absence of astrocyte. Pr has been calculated as the ratio of the number of successful post-synaptic responses to the number of pre-synaptic impulses (with a time-window of length 5 seconds).

3.2 Astrocyte-dependent Information Processing

In this subsection we show simulations associated with the biophysical model governed by equations (1) - (20) i.e., the astrocyte-dependent information processing. In Figure 6, we give an idea of the processes involved in the loop shown in Figure 2B. For the simulation of the scheme, shown in Figure 2B, we simultaneously solved equations (1) - (20). Of particular interest is the astrocyte-dependent feed-forward and feed-back paths making up a loop (Figure 2B). The same input signal was used in a feed-back manner into the loop. Using such a feed-forward and feed-back pathway an input signal can be amplified as per the cognitive process requiring strengthening of synapses. Ultimately such a process, may, lead to enhanced synaptic efficacy.

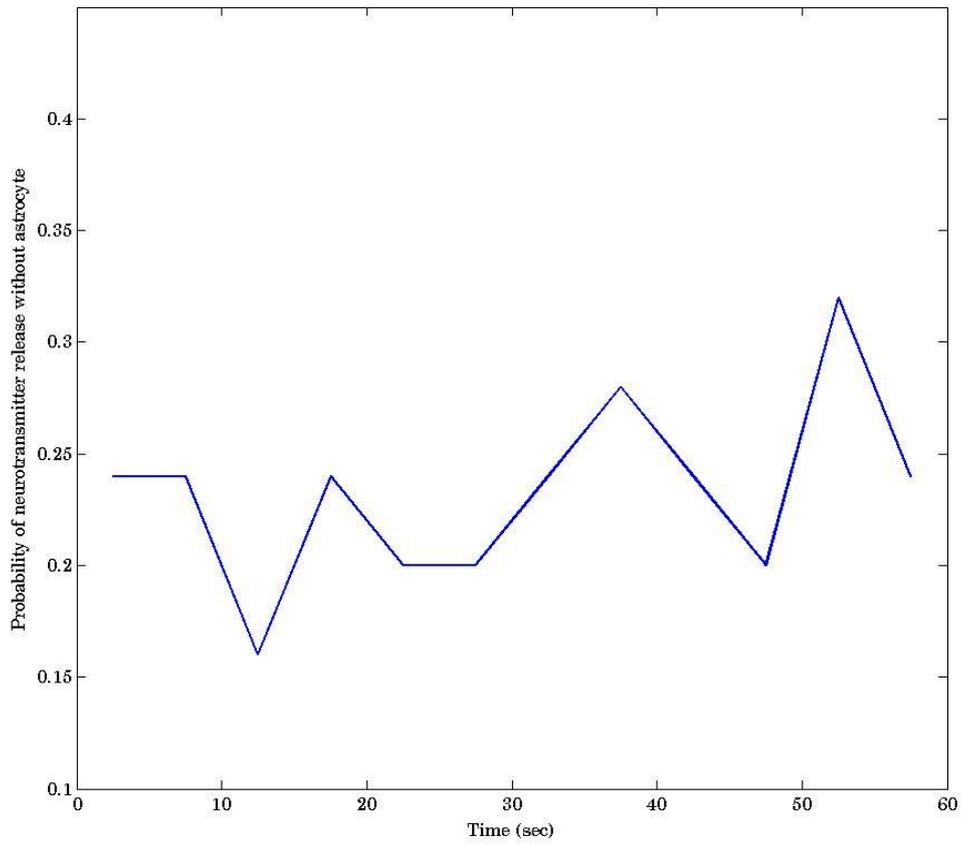

Figure 5. Probability of neurotransmitter release Pr, without incorporating the feedback loop due to astrocyte, is computed as the ratio of the number of successful post-synaptic response to the number of pre-synaptic stimulus (which was 5 Hz for the given simulation) within a time-window of length 5 seconds.

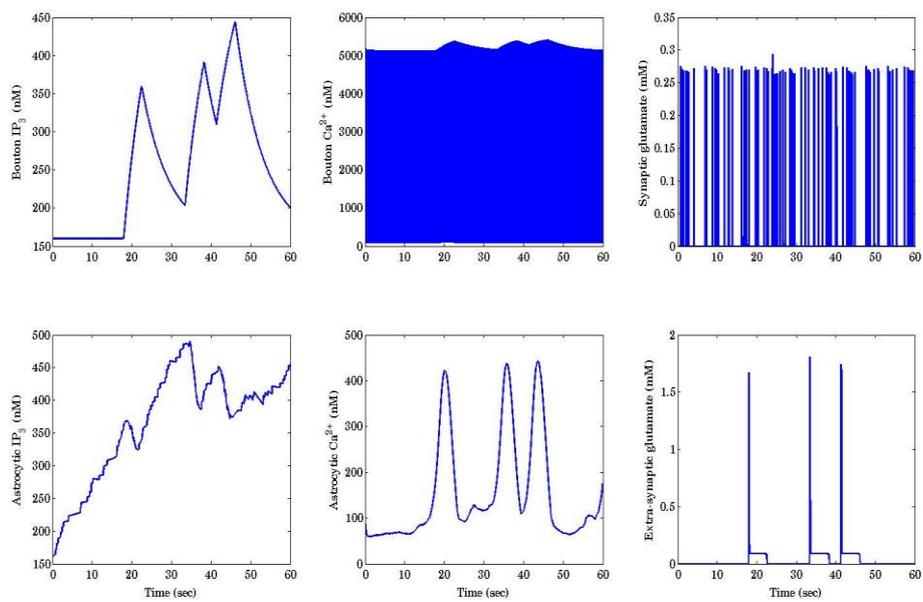

Figure 6. The major variables involved in astrocyte-dependent information processing. Here, input signal is same as in Figure 3A and is omitted. Output F is feeding back into the input A. A. Increased bouton IP$_3$ concentration in response to elevated extra-synaptic glutamate concentration (see F). B. Increased IP$_3$ concentration causes the IP$_3$R channels to open and leads to a transient enhancement in bouton [Ca$^{2+}$], due to influx of Ca$^{2+}$ from IP$_3$R (see Ca$^{2+}$ concentration after 20 seconds). C. Accumulated bouton [Ca$^{2+}$] leads to increased transients of glutamate concentration in the synaptic cleft. D. Transients of glutamate concentration set-off the production of astrocytic IP$_3$ concentration through an mGluR dependent pathway. E. Elevated astrocytic IP$_3$ concentration causes the IP$_3$R channels to open and initiates astrocytic Ca$^{2+}$ oscillations. F. Astrocytic Ca$^{2+}$ oscillations instigate the process of SLMV fusion, which is followed by a raised extra-synaptic glutamate concentration. This elevated extra-synaptic glutamate concentration forms the basis of bouton IP$_3$ production shown in A.

All the variables shown in Figure 6 are inter-dependent i.e., variation in one affects variation in others. When the bouton is fed with an input signal, it shows its response, in the form of increased cytosolic [Ca$^{2+}$] (see Figure 6B). This elevated [Ca$^{2+}$] exocytose glutamate in the synaptic cleft (see Figure 6C). After being exocytosed, synaptic glutamate can have either of the two fates (see Figure 2B). It can either bind with the post-synaptic AMPAR or it can bind with the mGluRs on the surface of the astrocyte. Once this glutamate binds with mGluR, it instigates the production of astrocytic IP$_3$ (see Figure 6D) through a G-protein link. During this glutamate spill-over process astrocytic IP$_3$ concentration goes on appreciating and gradually starts oscillating (notice after the 20 seconds mark Figure 6D). It can be observed from Figure 6D and Figure 6E that astrocytic Ca$^{2+}$ also starts oscillating as soon as astrocytic IP$_3$ starts oscillating. The biological significance and importance of IP$_3$ oscillation on Ca$^{2+}$ oscillation is not been fully understood though (De Pitta et al., 2009). This astrocytic Ca$^{2+}$ is known to exocytose SLMVs filled with glutamate once it crosses its threshold value of 196.69 nM (Parpura & Haydon, 2000). Similarly, in our model whenever astrocytic Ca$^{2+}$ crosses its threshold value it can spill glutamate (contained in SLMVs) in the extra-synaptic cleft (see Figure 6F). We have mathematically modeled this process of astrocytic glutamate release using equations (15)-(18). Extra-synaptic glutamate binds with extra-synaptic mGluR located on the surface of the bouton and initiates the production of bouton IP$_3$ (see Figure 6A) through a G-protein link. It is visible from Figure 6F and Figure 6A that bouton IP$_3$ production starts only when the astrocyte spills glutamate in the extra-synaptic cleft, reflecting the significance of extra-synaptic glutamate in the model. This bouton IP$_3$ is free to diffuse inside the cytosol and opens the IP$_3$R on the intracellular stores in a Ca$^{2+}$-dependent manner. Transient accumulation of Ca$^{2+}$ takes place as a result of opening up of IP$_3$R on the surface of the intracellular store (e.g., see Figure 6B at 20 seconds mark). Flow of Ca$^{2+}$ through these IP$_3$Rs is a slow process and is known to play a crucial role in modulating synaptic plasticity and spontaneous vesicle release (Emptage et al., 2001).

The synaptic vesicle exocytosis from bouton and SLMV release from astrocyte has been modeled using equations (7) - (9) and equations (15) - (17). Figure 7A and Figure 7B show the fraction of releasable and effective vesicles respectively during synaptic vesicle recycling process emulated using equations (7) - (9). Figure 7A and 7B are similar to the diagrams in Figure 4, except for the astrocyte-dependent pathway used here. The SLMV recycling process has been modeled using equation (17).

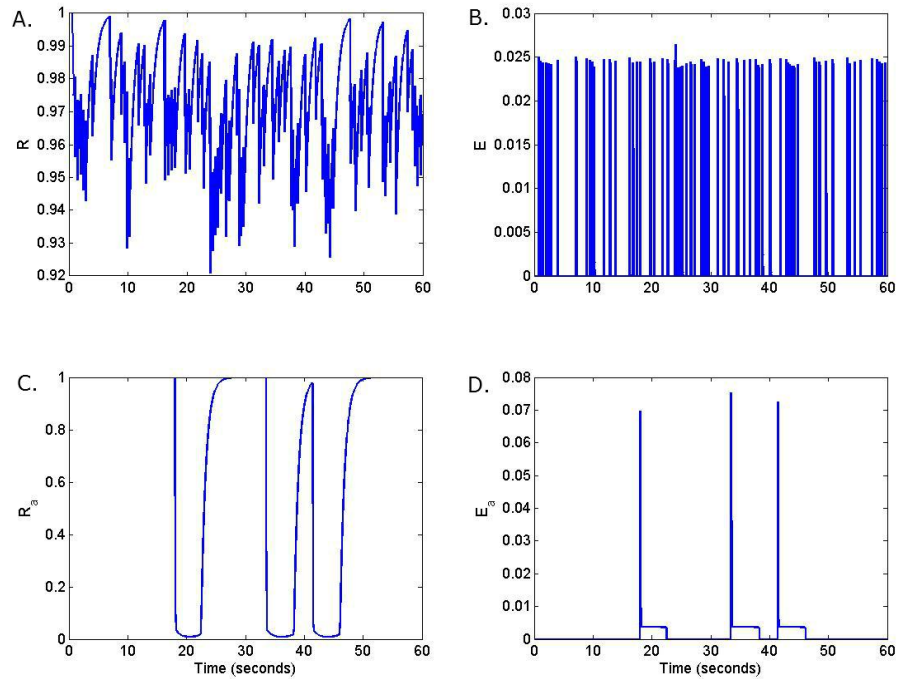

Figure 7. Fraction of releasable and effective vesicles, in astrocyte-dependent information processing, during an input AP of 5 Hz (see Figure 3A). A. Fraction of releasable vesicles inside the bouton. B. Fraction of effective vesicles in the synaptic cleft i.e., fraction of vesicles fused and residual vesicles in the synaptic cleft. C. Fraction of releasable SLMVs inside the astrocyte. D. Fraction of effective SLMVs in the extra-synaptic cleft i.e., fraction of SLMV fused and residual SLMV in the extra-synaptic cleft.

Figure 7C and Figure 7D show the fraction of releasable vesicles in astrocyte and effective vesicles in extra-synaptic cleft. It can be observed from Figure 7A that nearly 92% of the releasable (docked) vesicles have been used in astrocyte-dependent pathway. The fraction of effective vesicles in the synaptic cleft has also considerably gone-up (compare with Figure 4B). It is because of the transient increase in $Ca^{2+}$ concentration (see Figure 6B) which improves synaptic vesicle release probability (see Figure 7). In fact, the average vesicle release probability during this pathway was nearly 0.35, implying more than one out of three spikes are able to release a synaptic vesicle. It should be pointed out that the similar amount of enhancement in neurotransmitter release probability has been observed experimentally as well. Perea & Araque (2007) reported an increased Pr after astrocyte stimulation (from 0.24 to 0.33). We next show neurotransmitter release probability following the astrocyte-dependent pathway of information processing. A transient increase in neurotransmitter release probability can be observed from Figure 8 in close correlation with the astrocytic $Ca^{2+}$ concentration (see Figure 6E). The average neurotransmitter release probability under astrocyte-dependent pathway of information processing was 0.338 compared to 0.23 for astrocyte-independent pathway.

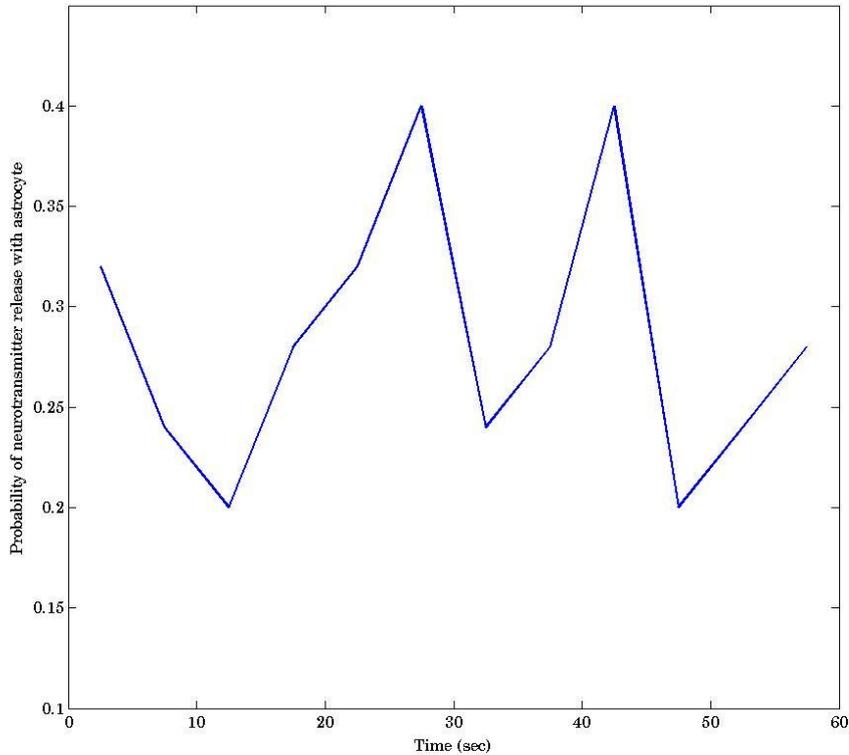

Figure 8. Probability of neurotransmitter release after incorporating the feed-forward and feed-back loop due to astrocyte, is computed as the ratio of the number of successful post-synaptic response to the number of pre-synaptic stimulus (which was 5 Hz for the given simulation) within a time-window of length 5 seconds.

3.3 Comparison between the two-forms of information processing

In this subsection, we have undertaken a comparative study between the two forms of information processing (see Figure 2A & 2B). We will discuss some of our findings keeping in mind the recent controversy regarding whether astrocytic $[Ca^{2+}]$ contributes in synaptic plasticity or not (e.g., Henneberger et al., 2010 vs. Agulhon et al., 2010).

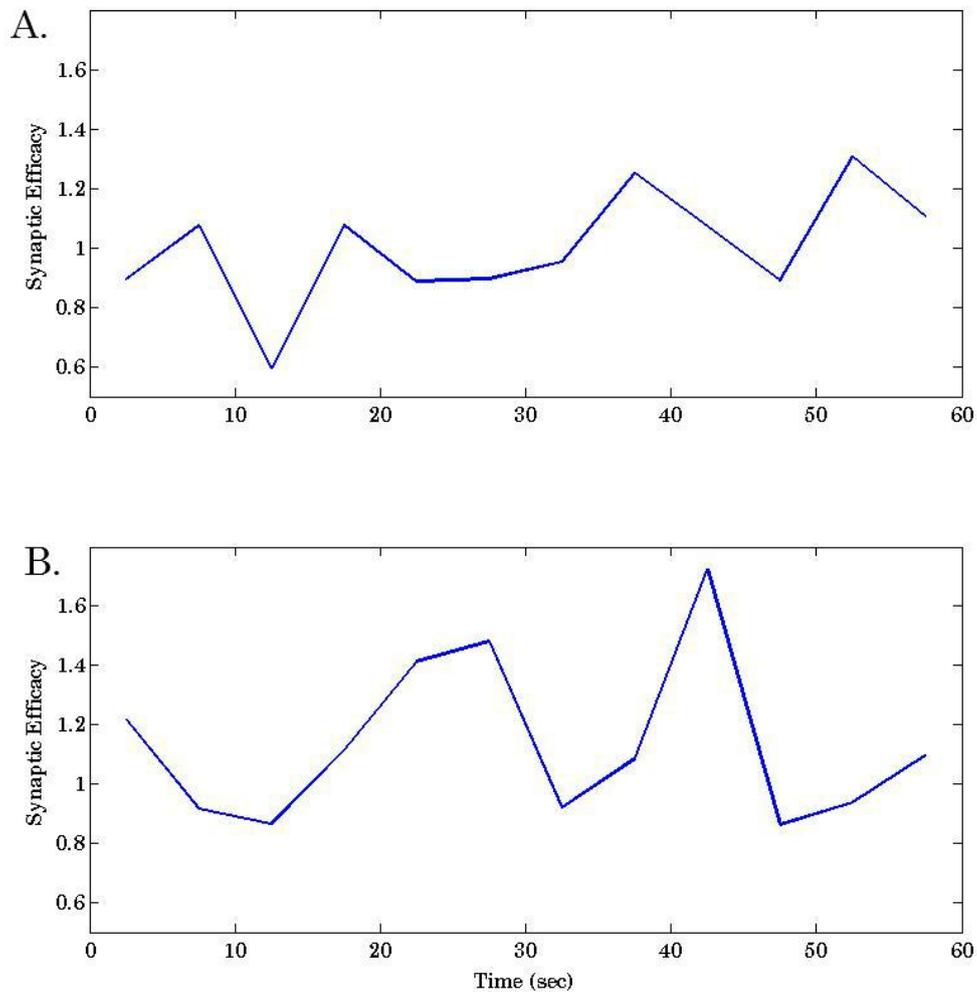

Figure 9. A comparison of the two modes of information processing (see Figure 2) in response to the same input signal of 5 Hz. Synaptic efficacy is calculated as the windowed-mean of post-synaptic responses including successes and failures where the window length has been taken to be 5 seconds for both figures. A. Output signal using astrocyte-independent information processing and B. Output signal using astrocyte-dependent information processing.

Using their experimental setup Perea & Araque (2007) demonstrated an increase in synaptic efficacy at single CA3–CA1 synapses during the phase of high astrocytic $[Ca^{2+}]$ (see Figure 1F of Perea & Araque, 2007). They stimulated the pre-synaptic neuron and simultaneously increased the astrocytic $[Ca^{2+}]$ through different pathways, e.g., purinergic receptors (P2Y-R), and recorded the EPSCs. They used caged $Ca^{2+}$ and used UV-flash to artificially increase astrocytic $[Ca^{2+}]$. In contrast, in our mathematical model, we allow an activity-dependent increase in astrocytic $IP_3$ following an AP. As a measure of change in synaptic strength i.e., synaptic efficacy, Perea & Araque (2007) demonstrated an increase in mean EPSC amplitude when astrocyte was stimulated. We measured the mean EPSC after every 5 sec. In Figure 9B, the mean EPSCs have been measured relative to the mean EPSC during first 20 seconds, because it is the phase during which astrocytic [$Ca^{2+}$] has not exceeded its threshold (see Figure 6E). In Figure 9A, the mean EPSCs have been measured relative to their overall mean. The impact of astrocytic response is clearly visible when we look at Figure 9A and 9B. In astrocyte-independent information flow, there is not much deviation ($\pm$ 20%) from its mean value, while in astrocyte-dependent information flow there is a transient increase of nearly 80%. This increase is subsequent to the rise in astrocytic $Ca^{2+}$ (see Figure 6E) and has decay time constant (the time necessary to reach 1/e of its initial magnitude (Fisher et al., 1997)) of nearly 10s. This increase in synaptic efficacy falls under short-term-enhancement, in particular augmentation, given the classification in Koch (1999, p – 311).

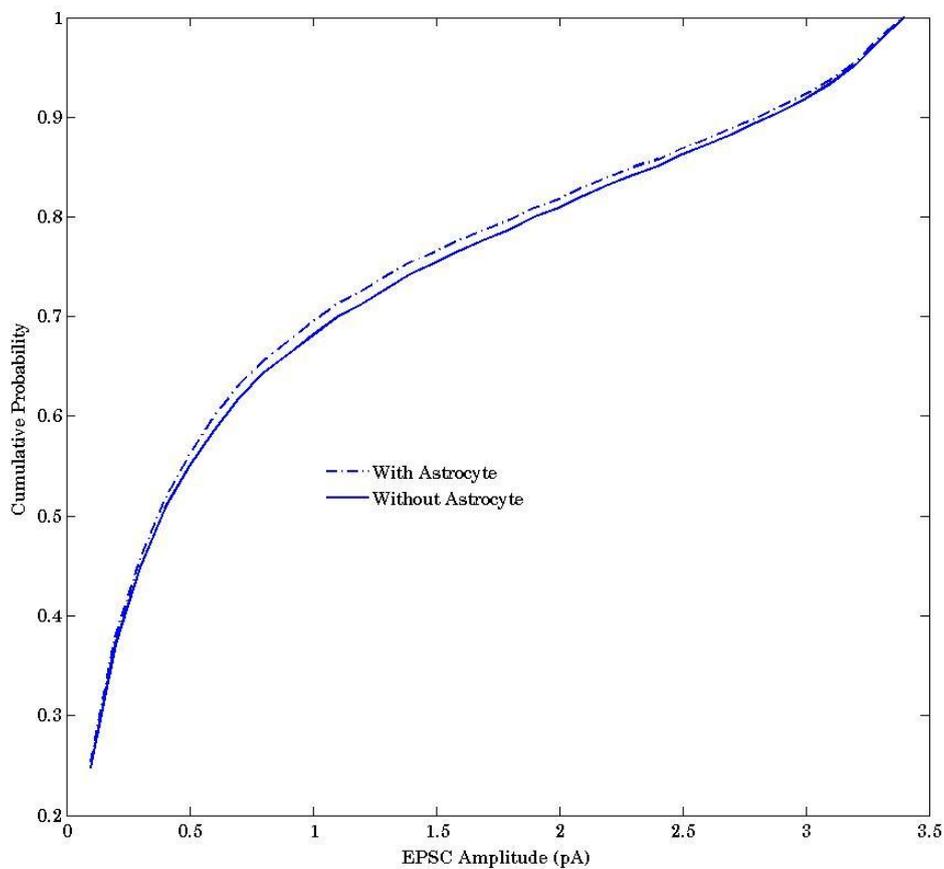

Figure 10. Cumulative probability of EPSC amplitude in response to an input signal of 5 Hz. Astrocyte-dependent curve shifts upwards implying an increased probability of having EPSC amplitude between 0.5 to 2.5 pA.

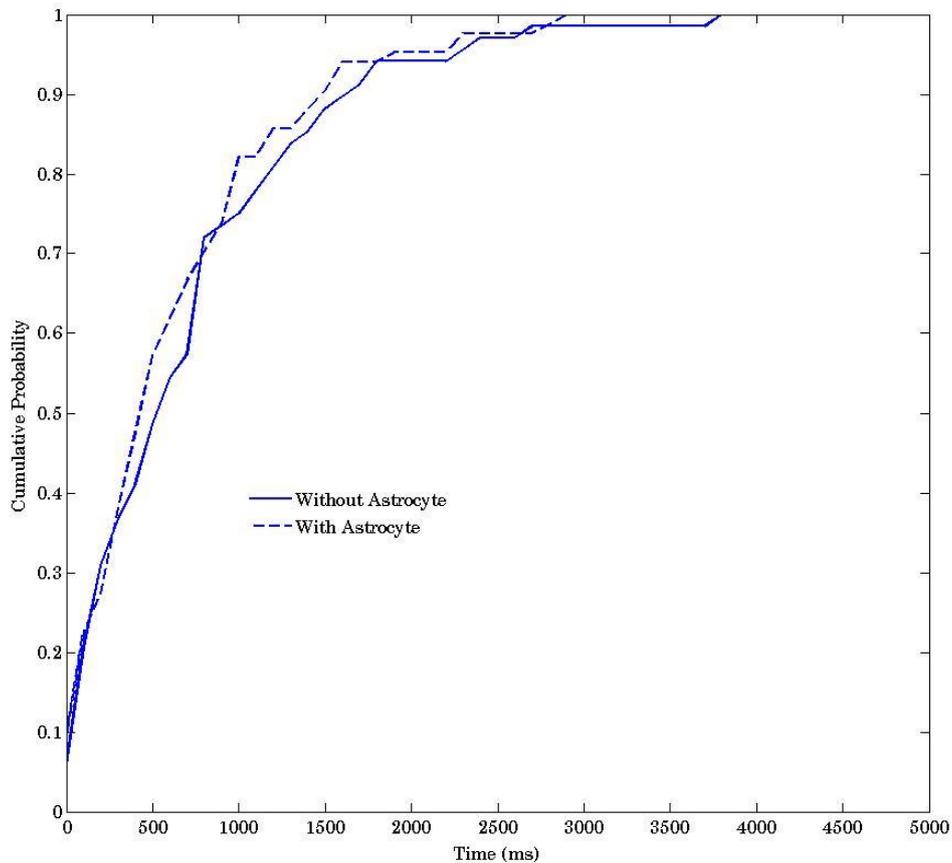

Figure 11. Cumulative probability distribution of inter-arrival time of EPSP for astrocyte-dependent and astrocyte-independent information processing. The distribution associated with astrocyte-dependent process shifts radically to the left suggesting reduced inter-arrival time due to enhanced synaptic efficacy.

Perea & Araque (2007) also demonstrated an increase in cumulative probability of EPSC amplitude before (astrocyte-independent) and during (astrocyte-dependent) astrocyte stimulation (see Figure 1E of Perea & Araque, 2007). Similar to their experimental observations, we also observed an increase in probability of EPSC amplitude (see Figure 10). This implies that there are more chances of having EPSC amplitude between 0.5 to 2.5 pA when astrocyte is present. Apart from an input signal of 5 Hz we also tested cumulative probability for an input signal of 2 Hz and 10 Hz. We observed that the astrocyte mediated potentiation (for an input signal of 2 Hz) of synaptic efficacy becomes more prominent as demonstrated by a significant increase in cumulative probability of observing EPSC amplitudes between 1.5 to 4.5 pA (see Figure 1 of the supplementary online material), similar to Figure 10 here. On the other hand, the contribution of astrocyte mediated potentiation (for an input signal of 10 Hz) of synaptic efficacy becomes less prominent or insignificant when compared with synaptic efficacy following astrocyte-independent pathway (see Figure 2 of the supplementary online material). The decrease in astrocyte mediated synaptic potentiation observed with an increase in the frequency of input signal might be due to the fact that our model has been calibrated for the experiments of Perea & Araque (2007) where they applied mild pre-synaptic neuron stimulation.

A more comprehensive way of demonstrating synaptic enhancement will be to show that we have more number of post-synaptic spikes under astrocyte-dependent processing than astrocyte-independent processing. In Figure 11, we show cumulative probability distribution for inter-arrival time of post-synaptic potentials. Cumulative probability graph tells us the probability of having a post-synaptic firing in a time-interval of length *x* ms (where *x* is an arbitrary point on the abscissa in Figure 11). Obviously the probability of having a post-synaptic spiking will increase as we increase the length of the time-interval (see Figure 11 where after 4000 ms mark cumulative probability is 1 under both forms of information processing). It is apparent from the figure that the probability of having post-synaptic spiking in short time-intervals has greatly increased in presence of astrocyte (see Figure 11).

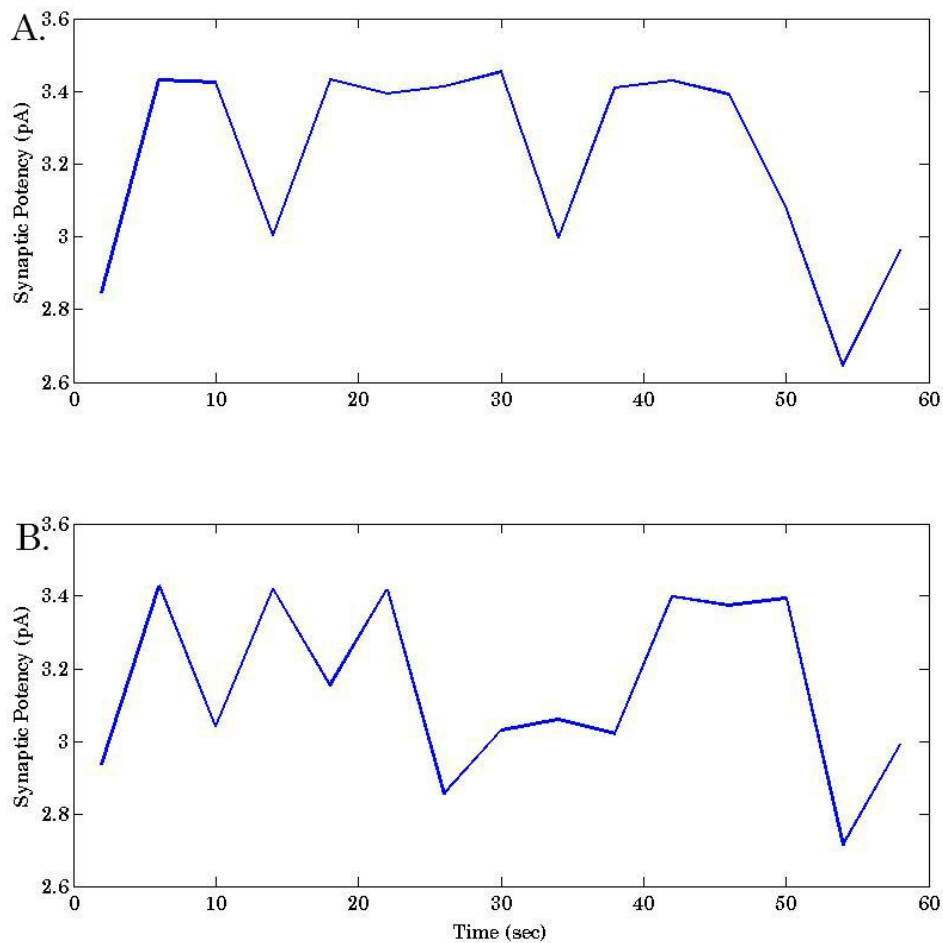

Figure 12. Synaptic potency under both forms of information processing (i.e., astrocyte-independent & astrocyte-dependent). Synaptic potency is given as a measure of mean EPSC, calculated over a time-window of 4-sec, excluding failures. Synaptic potency is unchanged in both cases which has also been observed in recent experiments (see Figure 1 of Perea & Araque, 2007); A. mean = -3.21 pA, std = 0.27 pA; B. mean = -3.11 pA, std = 0.24 pA. The two-sample paired t-test helps establish the previous statement ($p = 0.4475$).

During this type of astrocyte-induced plasticity, it is known that synaptic potency remains unchanged (Perea & Araque, 2007). Synaptic potency is given as a measure of mean post-synaptic response, excluding failures. We calculated the mean of each successful post-synaptic response in a time-window of 4 sec. It can be observed from Figure 12 that there is no apparent difference in synaptic potency under both forms of information processing. This observation was also confirmed statistically using a two-sample student's t-test. Synaptic potencies were assumed to be independent normally distributed random samples. It was tested that both the samples are with equal mean and equal but unknown variances (*null hypothesis*), against the alternative that the means are not equal with 5% significance level. The result returned a *p*-value of 0.4475 indicating a failure to reject null hypothesis.

Using our simulation, we found that, all these measures (like synaptic efficacy, inter-arrival time) which are used to demonstrate and establish the effect of astrocyte-dependent pathway over synaptic plasticity depend primarily on two parameters i) size of readily releasable pools of SLMVs in astrocyte and ii) rate of $IP_3$ production due to pre-synaptic mGluRs. The size of readily releasable pools has recently been determined using astrocyte cultures (Malarkey & Parpura, 2011). Here we show change in neurotransmitter release probability for a readily releasable pool ½ (see Figure 13A) and 1½ (see Figure 13C) in size of readily releasable SLMV pool determined experimentally (see Figure 13B). Computer simulations shown in Figure 13A–13C reveal the effect of different sizes of readily releasable pool of SLMVs. It is apparent that for a readily releasable pool of size 6 (i.e., containing 6 SLMVs) astrocytes do not contribute to enhance pre-synaptic neurotransmitter release probability. In fact, the average neurotransmitter release probability for readily releasable pool of size 6 was 0.25, which is similar to the average neurotransmitter release probability without astrocyte.

Figure 13B is the simulation of the model for default set of parameters listed in Table 2–Table 7. It is apparent from the figure that increase in neurotransmitter release probability is preceded with increase in astrocytic $Ca^{2+}$ concentration. In Figure 13C we again show neurotransmitter release probability but for an increased size of readily releasable SLMV pool. The effect of increased pool size is apparent from Figure 13C. The average neurotransmitter release probability in this case was 0.35. It should be noted that coherence between astrocytic $[Ca^{2+}]$ (see Figure 13D–13F) and neurotransmitter release probability (see Figure 13A–13C) is absent only for $n_a^v = 6$ (compare Figure 13A and Figure 13D) which highlights a possible biological condition under which astrocyte does not contribute to synaptic plasticity. The average neurotransmitter release probability in three simulations was 0.25, 0.33 and 0.35. There is no considerable difference between the experimentally determined pool size and a pool size of 18 (i.e. containing 18 SLMVs). However there was considerable difference in the maximum extra-synaptic glutamate concentration when latter compared with former (2.56 mM against 1.8 mM; data not shown). It is because of the negative cooperativity of mGluRs in response to extra-synaptic glutamate binding which ensures robust response to lower concentration of glutamate and also ensures insensitivity to higher concentration of glutamate. Thus, extra-synaptic glutamate is necessary for astrocyte mediated synaptic potentiation but with limited influence. A more potent contributor to astrocyte mediated synaptic potentiation is the $IP_3$ production rate by pre-synaptic mGluRs.

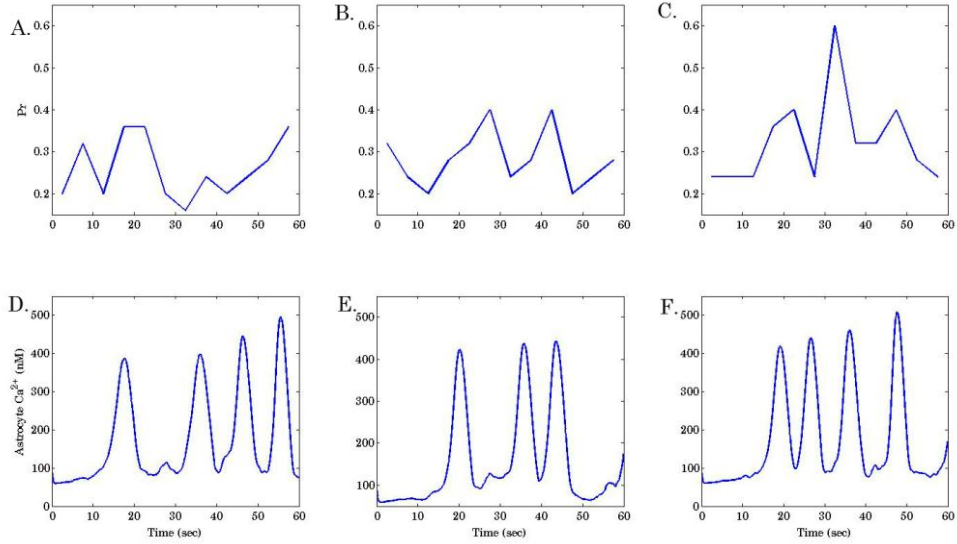

Figure 13. Neurotransmitter release probability in response to changing availability of readily releasable SLMV pool inside astrocyte. A. Neurotransmitter release probability for a readily releasable SLMV pool of size 6. B. Neurotransmitter release probability for a readily releasable SLMV pool of size 12. C. Neurotransmitter release probability for a readily releasable SLMV pool of size 18. D – F Astrocytic $Ca^{2+}$ concentration corresponding to the three simulations shown from A – C.

The maximum rate of $IP_3$ production $v_g$, by pre-synaptic mGluRs can be expressed in terms of surface density $\rho_{mGluR}$ of mGluR, let the surface area of bouton exposed to extra-synaptic glutamate released by astrocyte be $S$, the Avogadro Number $N_A$, the volume of bouton $V_{btn}$, and the production rate of $IP_3$ molecule per receptor $r_p$. Then

$$v_g = \frac{r_p \rho_{mGluR} S}{V_{btn} N_A}. \tag{22}$$

Nadkarni & Jung (2007) estimated the maximum rate of $IP_3$ production to be 0.062 nM ms$^{-1}$ or $\dfrac{0.062 \times 10^{-9} \times 6.023 \times 10^{23}}{10^{-3}}$ molecules/m$^3$ms or $0.373 \times 10^{17}$ molecules/ms per unit volume by mGluRs on the surface of the astrocyte. Such an estimate of $IP_3$ production rate is not known at boutons of CA3 pyramidal neurons. Thus, we used the IP$_3$ production rate by mGluRs on the pre-synaptic bouton to be same as that determined by Nadkarni & Jung (2007) i.e., 0.062 nM per ms. Hence, for an average bouton (of volume 0.13 µm$^3$, Koester & Sakmann (2000)) at hippocampal CA3-CA1 synapse the production rate will be $0.373 \times 10^{17} \times 0.13 \times 10^{-18} \approx 0.0048$ molecules/ms. If we assume $(2 \times \pi \times 0.31 \times 0.0028)$ i.e. $\approx 0.0055$ µm$^2$ (0.31 µm is the radius of bouton and 0.0028 µm is the strip of bouton exposed to extra-synaptic glutamate) of bouton is exposed to glutamate released in the extra-synaptic cleft by the astrocyte. Also if we assume that receptors produce 1 IP$_3$ molecule per ms, then the receptor density on relevant surface of the bouton is $\approx 0.87$ per µm$^2$. This assessment is in conformity with the experiments as receptor density at synapses is estimated to be between 200 – 2000 / µm$^2$ (Holmes, 1995) and extra-synaptic receptor density is known to be 230 times less than the receptor density at the synapse (Nusser et al., 1995). The exact density of extra-synaptic mGluRs on CA3 pyramidal neurons is not known. Hence, we simulated the model for a range of possible IP$_3$ production rates (see Figure 14A – D). The average neurotransmitter release probability for $v_g = 0.05$ nM ms$^{-1}$ is nearly equal to astrocyte-independent pathway of information processing (Pr = 0.24 against Pr = 0.23). But as we increased the value of $v_g$ the effect of astrocyte over synaptic plasticity became more prominent. The average neurotransmitter release probability for $v_g = 0.1$ nM ms$^{-1}$ and $v_g = 0.2$ nM ms$^{-1}$ was 0.36 and 0.4 respectively. Please note that Figure 14B is same as Figure 13B and Figure 8, it has been shown for comparison purposes only. Our simulation reveals that $v_g$ is a critical parameter which can modulate the contribution of astrocyte induced synaptic plasticity.

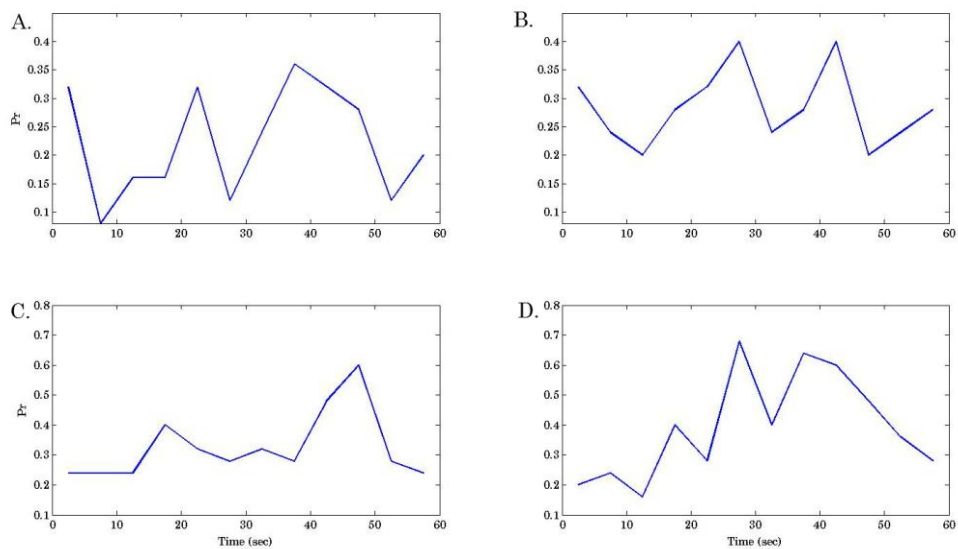

Figure 14. Plasticity of Neurotransmitter release probability in response to varying rate of IP$_3$ production by pre-synaptic group I mGluRs. A. Neurotransmitter release probability for an IP$_3$ production rate of 0.05 µM per sec. B. Neurotransmitter release probability for an IP$_3$ production rate

of 0.062 μM per sec. C. Neurotransmitter release probability for an IP$_3$ production rate of 0.1 μM per sec. D. Neurotransmitter release probability for an IP$_3$ production rate of 0.2 μM per sec. Please note the change in Y-axis bounds for C and D.

4. Conclusion and future directions

There is a debate regarding the mechanism and calcium dependence of gliotransmission and the role of gliotransmission in synaptic plasticity. Together they imply that the effect of astrocytic calcium on synaptic plasticity is a controversial issue. Here we have put together a number of phenomenological and biophysical models for the processes shown in Figure 2 to simulate the effects on synaptic strength with and without astrocytic Ca$^{2+}$. From the computational modeling point of view this is equivalent to controlling the effect of Ca$^{2+}$ in astrocytes by genetic engineering (Agulhon et al., 2010) and by calcium clamp (Henneberger et al., 2010) in order to study the effects of astrocytic Ca$^{2+}$ on synaptic plasticity. A better understanding, through varieties of approaches, of calcium dynamics, signaling and gliotransmitter release is necessary for settling down the aforementioned debate (Ben Achour et al., 2010). Here we have taken a computational approach, and concluded that the astrocytic Ca$^{2+}$ does contribute to the synaptic augmentation at the time scale of the order of seconds, for the given mathematical framework.

Here we presented a mathematical model which studies the effect of astrocyte over the hippocampal CA3–CA1 synaptic strength. It is found that given the pathway (Figure 2B), astrocyte plays a significant role in modulating synaptic information transfer. It might be possible that under physiological conditions, neurons also exhibit the two types of information processing: i) astrocyte-independent ii) astrocyte-dependent. Recent study performed by Di Castro et al. (2011) confirms that astrocytes are activated under physiological stimulation of neighboring synapses. It is suggested that neurons process information usually in astrocyte-independent manner unless there is a need to modify synaptic efficacy according to various plasticity events taking place in hippocampus (Navarrete & Araque, 2010; Panatier et al., 2011; Navarrete et al., 2012).

Using our computational model, we identified two important parameters (readily releasable pool size of SLMVs and maximum rate of IP$_3$ production rate) which affect astrocyte mediated synaptic potentiation at single CA3–CA1 synapse. Our simulations reveal a possible biological condition under which astrocyte Ca$^{2+}$ oscillations *do not* contribute to synaptic potentiation (see Figure 13A). It was found that maximum rate of IP$_3$ production rate ($v_g$) was a more potent modulator (of the two parameters) of astrocyte mediated synaptic potentiation. Using equation (22) and performing simple algebraic calculations we could predict mGluR density on relevant surface of CA3 pyramidal neuron bouton which is experimentally unknown at CA3–CA1 synapse but was in conformity with experiments from other synapses (Nusser et al., 1995; Holmes, 2000).

It should be pointed out that, it is not possible to conclude and assert that astrocyte induces a particular type of synaptic plasticity (e.g., augmentation) using only a temporal model, like the one proposed here, as synaptic plasticity depends on several spatial constraints. As a future direction, it is proposed to develop a spatio-temporal model to study the effects of spatial constraints, like release sites, $Ca^{2+}$ sources etc., over modulation of synaptic activity. It is also known that a single hippocampal astrocyte in CA1 region ensheaths around thousands of synapses (Schipke & Peters, 2009). Thus, it is possible for a single astrocyte to modulate signal processing at thousands of synapses simultaneously. It has also been shown experimentally that, astrocytes help to synchronize firings of neurons in CA1 region (Carmignoto & Fellin, 2006). Hence, it is worthy to study the effects of astrocytes over the networks of neurons. Porto−Pazos et al. (2011) recently performed a study where they highlighted the importance of artificial astrocytes in modulating an artificial neural network.

The present mathematical model is quite adaptable and can be easily extended to study longer and other forms of synaptic plasticity (Tewari & Majumdar, 2012). Another advantage of this model is that it can be extended to astrocytic microdomains, where it is difficult to experimentally manipulate calcium fluctuations. Simply increasing intracellular calcium is not sufficient for gliotransmitter release, as evident from conflicting results (Henneberger et al., 2010; Agulhon et al., 2010; Wenker, 2010). If calcium is required for transmitter release, then it may need to occur in specific microdomains (Wenker, 2010), which has been over-looked and needs examination using similar computational modeling approaches among others.

Acknowledgments

The work has been supported by the Department of Science and Technology, Government of India, grant no. SR/CSI/08/2009. Helpful suggestions from Vladimir Parpura are being thankfully acknowledged.